\documentclass[twocolumn]{aastex63}

\pdfoutput=1 
\UseRawInputEncoding
\usepackage{color}
\usepackage{natbib,graphicx,amsmath}
\def\ci {[C{\scriptsize I}]}
\def\cii {[C{\scriptsize II}]}
\def\ciisub {[C{\scriptscriptstyle II}]}
\def\ciimath {[C{\scriptstyle II}]}
\def\oi {[O{\scriptsize I}]}
\def\oisub {[O{\scriptscriptstyle I}]}

\def\nii {[N{\scriptsize II}]}
\def\niisub {[N{\scriptscriptstyle II}]}
\def\niimath {[N{\scriptstyle II}]}
\def\oiii {[O{\scriptsize III}]}

\def\hii {H{\scriptsize II}}

\def\hiimath {[H{\scriptstyle II}]}
\def\lsol {$\rm{L_{\odot}}$}
\def\msol {$\rm{M_{\odot}}$}
\def\lfir {$\rm{L_{FIR}}$}
\def\micron {$\rm{\mu m}$}

\shorttitle{ISM in SPT0346-52}
\shortauthors{Litke et al.}

\begin{document}

\title{Multi-Phase ISM in the $\MakeLowercase{z} = 5.7$ Hyperluminous Starburst SPT0346-52}

\author[0000-0002-4208-3532]{Katrina~C.~Litke} 
\affiliation{Steward Observatory, University of Arizona, 933 North Cherry Avenue, Tucson, AZ 85721, USA; \href{mailto:kclitke@email.arizona.edu}{kclitke@email.arizona.edu}}
\correspondingauthor{Katrina~C.~Litke}
\email{kclitke@email.arizona.edu}
\author[0000-0002-2367-1080]{Daniel~P.~Marrone}
\affiliation{Steward Observatory, University of Arizona, 933 North Cherry Avenue, Tucson, AZ 85721, USA; \href{mailto:kclitke@email.arizona.edu}{kclitke@email.arizona.edu}}
\author[0000-0002-6290-3198]{Manuel~Aravena}
\affiliation{N\'ucleo de Astronom\'ia, Facultad de Ingenier\'ia y Ciencias, Universidad Diego Portales, Av. Ej\'ercito 441, Santiago, Chile}
\author[0000-0002-3915-2015]{Matthieu~B\'ethermin}
\affiliation{Aix Marseille Univ., Centre National de la Recherche Scientifique, Laboratoire dÕAstrophysique de Marseille, Marseille, France}
\author{Scott~C.~Chapman}
\affiliation{Eureka Scientific, Inc. 2452 Delmer Street Suite 100, Oakland, CA 94602-3017}
\affiliation{Dalhousie University, Halifax, Nova Scotia, Canada}
\author[0000-0002-5823-0349]{Chenxing~Dong}, 
\affiliation{Department of Astronomy, University of Florida, Gainesville, FL 32611, USA}
\author[0000-0003-4073-3236]{Christopher~C.~Hayward}
\affiliation{Center for Computational Astrophysics, Flatiron Institute, 162 Fifth Avenue, New York, NY 10010, USA}
\author{Ryley~Hill}
\affiliation{Department of Physics and Astronomy, University of British Columbia, 6225 Agricultural Road, Vancouver, V6T 1Z1, Canada}
\author[0000-0002-5386-7076]{Sreevani~Jarugula}
\affiliation{Department of Astronomy and Department of Physics, University of Illinois, 1002 West Green St., Urbana, IL 61801}
\author[0000-0001-6919-1237]{Matthew~A.~Malkan}
\affiliation{Department of Physics and Astronomy, University of California, Los Angeles, CA 90095-1547, USA}
\author[0000-0002-7064-4309]{Desika~Narayanan}
\affiliation{Department of Astronomy, University of Florida, Gainesville, FL 32611, USA}
\affiliation{University of Florida Informatics Institute, 432 Newell Drive, CISE Bldg E251, Gainesville, FL 32611}
\affiliation{Cosmic Dawn Center (DAWN), DTU-Space, Technical University of Denmark, DK-2800 Kgs. Lyngby, Denmark; Niels Bohr Institute, University of Copenhagen, Juliane Maries vej 30, DK-2100 Copenhagen, Denmark}
\author[0000-0001-7477-1586]{Cassie~A.~Reuter}
\affiliation{Department of Astronomy and Department of Physics, University of Illinois, 1002 West Green St., Urbana, IL 61801}
\author[0000-0003-3256-5615]{Justin~S.~Spilker}
\affiliation{Department of Astronomy, University of Texas at Austin, 2515 Speedway Stop C1400, Austin, TX 78712, USA}
\author[0000-0002-3187-1648]{Nikolaus~Sulzenauer}
\affiliation{Max-Planck-Institut f\"{u}r Radioastronomie, Auf dem H\"{u}gel 69 D-53121 Bonn, Germany}
\author[0000-0001-7192-3871]{Joaquin~D.~Vieira}
\affiliation{Department of Astronomy and Department of Physics, University of Illinois, 1002 West Green St., Urbana, IL 61801}
\author[0000-0003-4678-3939]{Axel~Wei{\ss}}
\affiliation{Max-Planck-Institut f\"{u}r Radioastronomie, Auf dem H\"{u}gel 69 D-53121 Bonn, Germany}

\begin{abstract}
SPT0346-52 ($z=5.7$) is the most intensely star-forming galaxy discovered by the South Pole Telescope, with $\rm{\Sigma_{SFR}} \sim 4200\ $\msol\ yr$^{-1}$ kpc$^{-2}$.  
 In this paper, we expand on previous spatially-resolved studies, using ALMA observations of dust continuum, \nii 205\micron, \cii 158\micron, \oi 146\micron, and undetected \nii 122\micron\  and \oi 63\micron\ emission to study the multi-phase interstellar medium (ISM) in SPT0346-52.  We use pixelated, visibility-based lens modeling to reconstruct the source-plane emission.  We also model the source-plane emission using the photoionization code \textsc{cloudy} and find a supersolar metallicity system.
 We calculate $T_{dust} = 48.3\ \rm{K}$ and $\lambda_{peak} = 80$\micron, and see line deficits in all five lines. 
 The ionized gas is less dense than comparable galaxies, with $n_e < 32\ \rm{cm^{-3}}$, while $\sim 20\%$ of the \cii 158 emission originates from the ionized phase of the ISM.  
   We also calculate the masses of several phases of the ISM.
We find that molecular gas dominates the mass of the ISM in SPT0346-52, with the molecular gas mass $\sim 4\times$ higher than the neutral atomic gas mass and $\sim 100\times$ higher than the ionized gas mass.

\end{abstract}

\keywords{High-redshift galaxies, Interstellar medium, Starburst galaxies}

\section{Introduction}

The interstellar medium (ISM) of high-redshift galaxies is difficult to study directly due to cosmological dimming and angular resolution limitations. Observations of rest-frame optical and ultraviolet wavelengths also suffer from significant dust extinction, and some phases of the ISM lack suitable tracers at these wavelengths. However, for very distant objects, far-infrared (FIR) continuum and line emission that is normally obscured by the Earth's atmosphere redshifts into the submillimeter window. The angular resolution and sensitivity afforded by the Atacama Large Millimeter/submillimeter Array (ALMA) is providing new opportunities to explore the physical conditions in early galaxies through their rest-frame FIR emission.

Many recent high-redshift studies \citep[e.g.,][]{gullberg2015,lefevre2020,bethermin2020} have focused on the the 158\micron\ fine structure line of singly-ionized carbon (hereafter, \cii 158\micron) because it is one of the brightest cooling lines of the ISM \citep{hollenbach1991}.
However, this line can be difficult to interpret because it can originate from both ionized gas and neutral gas in photo-dissociation regions (PDRs).
The 122 and 205 \micron\ lines of ionized nitrogen (\nii) arise from the ionized phase of the ISM because nitrogen has a higher ionization energy than hydrogen.
Since \nii 122\micron\ and \nii 205\micron\ trace the ionized ISM, comparing \cii 158\micron\  to \nii 205\micron\  emission makes it possible to determine what fraction of the \cii 158\micron\ emission originates from PDRs, with values typically in the $60-90\%$ range \citep[e.g.,][]{pavesi2016,diazsantos2017,herreracamus2018a,cormier2019}.
\oi 63\micron\ and \oi 146\micron, on the other hand, originate from warm, neutral gas \citep{tielens1985,hollenbach1991}.
Where there is more \oi 146\micron\  emission compared to the \cii 158\micron\ emission, we would expect more dense, neutral gas in those regions \citep{debreuck2019}.

Recently, (mostly) spatially unresolved multi-line surveys of high-z galaxies, including \nii 205\micron, \cii 158\micron, \oi 146\micron, and \nii 122\micron, have been conducted in individual systems.
\cite{novak2019} and \cite{debreuck2019} found highly enriched ISM, with approximately solar metallicities, in J1342+0928 and SPT0418-47.
\cite{debreuck2019} and \cite{lee2021} also found evidence for a dense gas-dominated ISM using the ratio of \oi 146\micron\ to \cii 158\micron\ in the first detections of \oi 146\micron\ at $z>1$.
\cite{rybak2020} also recently published the first \oi 63\micron\ detection at $z>3$ in a dusty galaxy at $z\sim 6$ and determined that \oi 63\micron\ was the main neutral gas coolant in G09.83808.

In this paper, we focus on the $z = 5.656$ gravitationally lensed dusty star-forming galaxy (DSFG) SPT-S J034640-5204.9 (hereafter SPT0346-52) \citep{weiss2013,vieira2013}.
SPT0346-52 is the most intensely star-forming galaxy from the 2500 deg$^2$ South Pole Telescope survey \citep[SPT;][]{vieira2010,carlstrom2011,everett2020}, with apparent \lfir$ = 1.1\times 10^{14}$\lsol \citep{spilker2015,reuter2020} and intrinsic star formation rate density $\rm{\Sigma_{SFR}} = 4200\ $\msol yr$^{-1}$ kpc$^{-2}$ \citep{hezaveh2013,ma2015,ma2016,spilker2015}, where
\lfir\ is the emission from $42.5-122.5$\micron\ \citep{helou1988}.
Based on \emph{Chandra} observations, \cite{ma2016} determined that the high \lfir\ is dominated by star formation with negligible contribution from an active galactic nucleus (AGN).

\cite{litke2019} performed pixelated, interferometric lens modeling of \cii 158\micron\ emission in SPT0346-52.  The gas in SPT0346-52 was found to be globally unstable, with Toomre Q instability parameters $\ll 1$ throughout the system.  In addition, they found two components separated by $\sim 1\ \rm{kpc}$ and $\sim 500\ \rm{km/s}$ that appear to be merging, which is likely driving the intense star formation in SPT0346-52.
More recently, \cite{jones2019} have suggested that a rotating disk galaxy is a better
explanation for a water absorption line.

In this paper, we extend the work of \cite{litke2019}, expanding their \cii 158\micron\  analysis to a survey of fine-structure lines.  Using \nii 205\micron, \cii 158\micron, \oi 146\micron, \nii 122\micron, \oi 63\micron, and the underlying dust continuum emission, we conduct a multi-phase study of the ISM in SPT0346-52.
This represents one of the first multi-line, spatially resolved studies of the ISM at high-z.

We describe ALMA observations of the five fine-structure lines in Section \ref{sec:obs}.
The lensing reconstruction process and results are discussed in Section \ref{sec:lens}.
In Section \ref{sec:cloudy}, we describe the \textsc{cloudy} modeling of the ISM in SPT0346-52.
We describe the results and various line and continuum diagnostics in Section \ref{sec:disc}, and summarize the results in Section \ref{sec:conc}.
We adopt the cosmology of \cite{planck} ($\Omega_m=0.309$, $\Omega_{\Lambda} = 0.691$, and $H_0 = 67.7\ \rm{km/s}$).  At $z=5.656$, 1'' = 6.035 kpc.

\section{ALMA Observations}
\label{sec:obs}

SPT0346-52 was observed in ALMA Bands 6, 7, and 9 from September 2014 through September 2018 (project IDs: 2013.1.01231, 2015.1.01580, 2016.1.01565; PI:  Marrone).  \cii 158\micron, \nii 122\micron, and \oi 63\micron\ were all observed on multiple dates at different resolutions, while \nii 205\micron\ and \oi 146\micron\ were each observed once.  Details of these observations, including dates, observing frequencies, flux and phase calibrators, and resolutions, are listed in Table \ref{table:obs}.

\begin{deluxetable*}{llccccccccc}
\centering
\tablecaption{ALMA Observations of SPT0346-52}
\tablenum{1}
\label{table:obs}
\tablehead{\colhead{Line} & \colhead{Date} & \colhead{Frequency$^{\mathrm a}$} & \colhead{\# Ant.} & \colhead{Resolution} & \colhead{Flux} & \colhead{Phase} & \colhead{PWV$^\mathrm b$} & \colhead{$\mathrm t_{\mathrm{int}}^\mathrm c$} & \colhead{Noise Level$^{\mathrm d}$} & \colhead{Project ID$^{\mathrm e}$} \\
\colhead{} & \colhead{} & \colhead{(GHz)} & \colhead{} & \colhead{(arcsec)} & \colhead{Calibrator} & \colhead{Calibrator} & \colhead{(mm)} & \colhead{(min)} & \colhead{(mJy/beam)} & \colhead{}}
\startdata
\textbf{Band 6} & & & & & & & & & \\
\nii 205\micron & 2015 Aug 30 & 227.518 & 35 & $0.19\times0.25$ & J0334-4008 & J0334-4008 & 1.4 & 44.3 & 0.10 & 2013.1.01231 \\
\hline
\textbf{Band 7} & & & & & & & & \\
\cii 158\micron  & 2014 Sep 2 & 291.533 & 34 & $0.22\times0.26$ & J0334-4008 & J0334-4008 & 0.9 & 5.3  & 0.25 & 2013.1.01231 \\
      & 2015 Jun 28 & 291.536 & 41 & $0.15\times0.17$ & J0334-4008 & J0334-4008 & 1.3 & 5.2  & 0.23 &              \\
\oi 146\micron  & 2014 Sep 2 & 304.136 & 34 & $0.22\times0.27$ & J2357-5311 & J0334-4008 & 0.9 & 23.8 & 0.20 &             \\
\nii 122\micron & 2016 Jun 30 & 364.434 & 40 & $0.28\times0.35$ & Ceres      & J0253-5441 & 0.6 & 24.6 & 0.21 & 2015.1.01580 \\
      & 2018 Aug 28 & 364.431 & 48  & $0.29\times0.37$     & J0519-4546 & J0253-5441 & 0.3 & 35.3 &   0.27   & 2016.1.01565 \\
\hline
\textbf{Band 9} & & & & & & & & \\
\oi 63\micron  & 2018 Aug 17 & 711.038 & 46 & $0.23\times0.33$     & J0522-3627 & J0210-5101 & 0.4 & 19.2 &   1.7   & 2016.1.01565 \\
      & 2018 Aug 19 & 711.038 & 45 & $0.23\times0.28$     & J0522-3627 & J0210-5101 & 0.4 & 19.2 &   1.5   &              \\
      & 2018 Sep 03 & 711.034 & 46 & $0.15\times0.19$     & J0522-3627 & J0210-5101 & 0.4 & 18.5 &   1.2   &              \\
      & 2018 Sep 03 & 711.034 & 46 & $0.16\times0.20$     & J0522-3627 & J0210-5101 & 0.3 & 19.2 &    1.4  &                
\enddata
\tablenotetext{a}{First local oscillator frequency}
\tablenotetext{b}{Precipitable water vapor at zenith}
\tablenotetext{c}{On-source integration time}
\tablenotetext{d}{Root-mean-square noise level in continuum image}
\tablenotetext{e}{PI: Marrone}
\end{deluxetable*}

The data were processed using various pipeline versions of the Common Astronomy Software Applications package \citep[CASA;][]{casa,petry2012}.
\cii 158\micron, \nii 205\micron, and \oi 146\micron\ were all processed using CASA pipeline version 4.2.2.
\nii 122\micron\ was processed with CASA pipeline version 4.7.1, and \oi 63\micron\ was processed with CASA pipeline version 5.4.0.
These were the accepted pipeline versions for the cycles in which each dataset was observed.

\begin{figure*}
\begin{center}
\includegraphics[width=\textwidth]{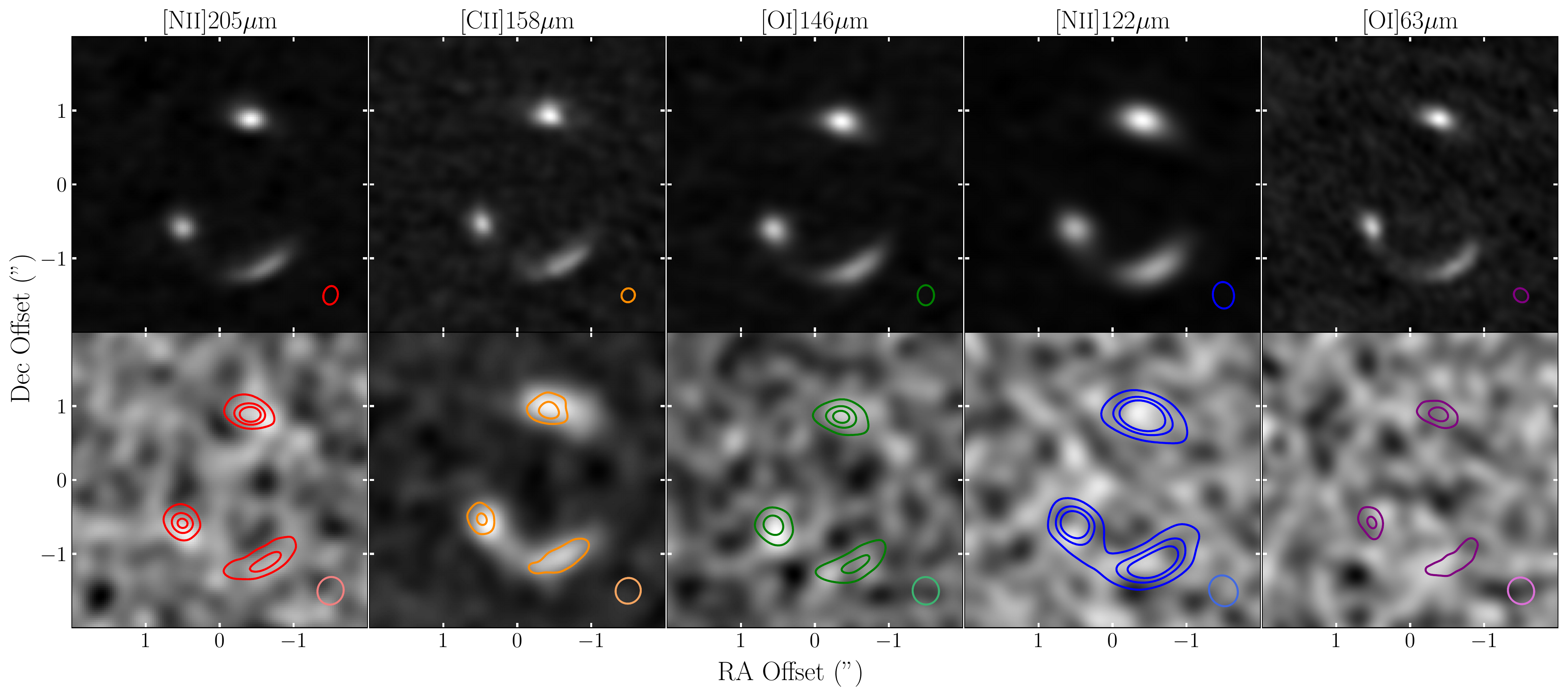}
\caption{Observed emission from SPT0346-52. Top:  Continuum images.  Bottom:  Line images with continuum contours.  Continuum images were untapered and used Briggs weighting (robust=0.5).  Line Images were made integrating from -300 to +300 km/s with Briggs weighting (robust=0.5) and tapered to $300k\lambda$.
The contours represent the observed continuum emission at 10, 30, and 50$\sigma$.   From left to right: 205\micron, 158\micron, 146\micron, 122\micron, and 63\micron.
\label{fig:obscont}}
\end{center}
\end{figure*}

Continuum images at all five frequencies were created with the task \textsc{tclean} in CASA version 5.4.0, using Briggs weighting (robust=0.5) and the \textsc{auto-multithresh} masking option.  The continuum images are shown in the top row of Figure \ref{fig:obscont} and as contours in the bottom row.
For the line emission, the continuum was subtracted from the line cube using the CASA task \textsc{uvcontsub} with a first order polynomial representing the continuum.

The line emission was imaged in the same manner as the continuum, but tapered to $300k\lambda$ and integrated from -300 to +300 km/s. (Figure \ref{fig:obscont}, bottom row).  This tapering was also used in the lensing reconstructions.

\begin{figure*}
\begin{center}
\includegraphics[width=0.32\linewidth]{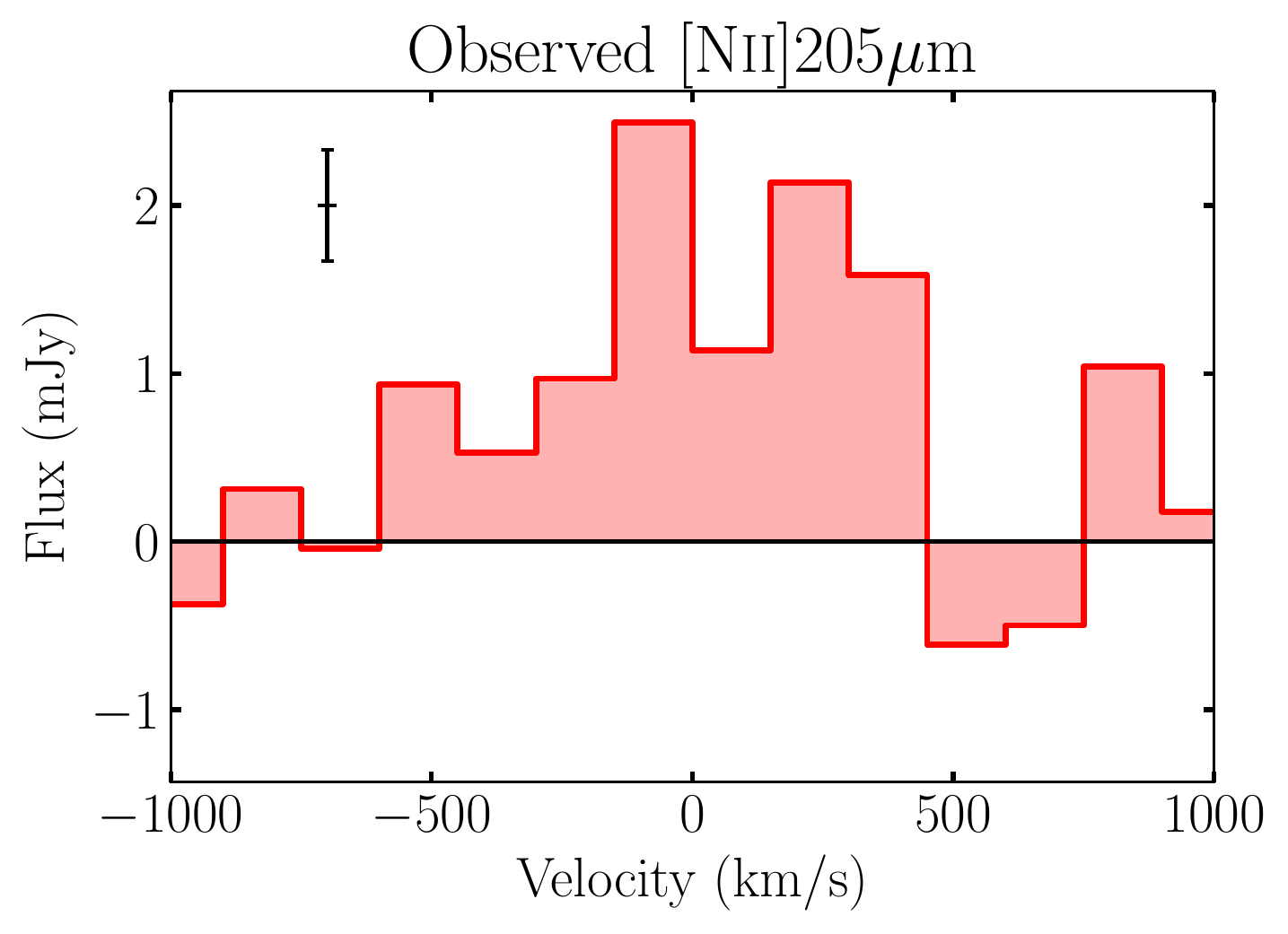}
\includegraphics[width=0.32\linewidth]{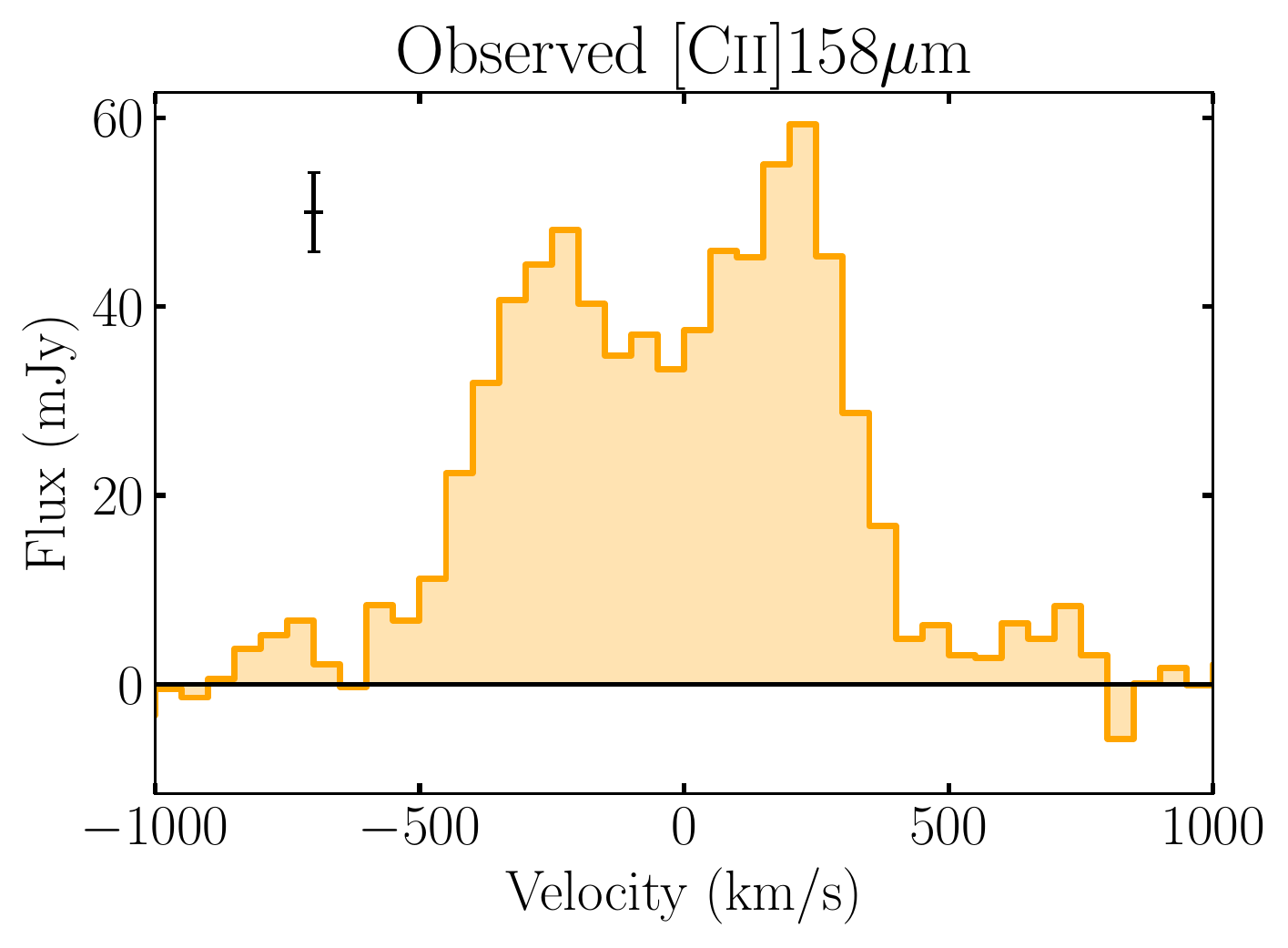}
\includegraphics[width=0.32\linewidth]{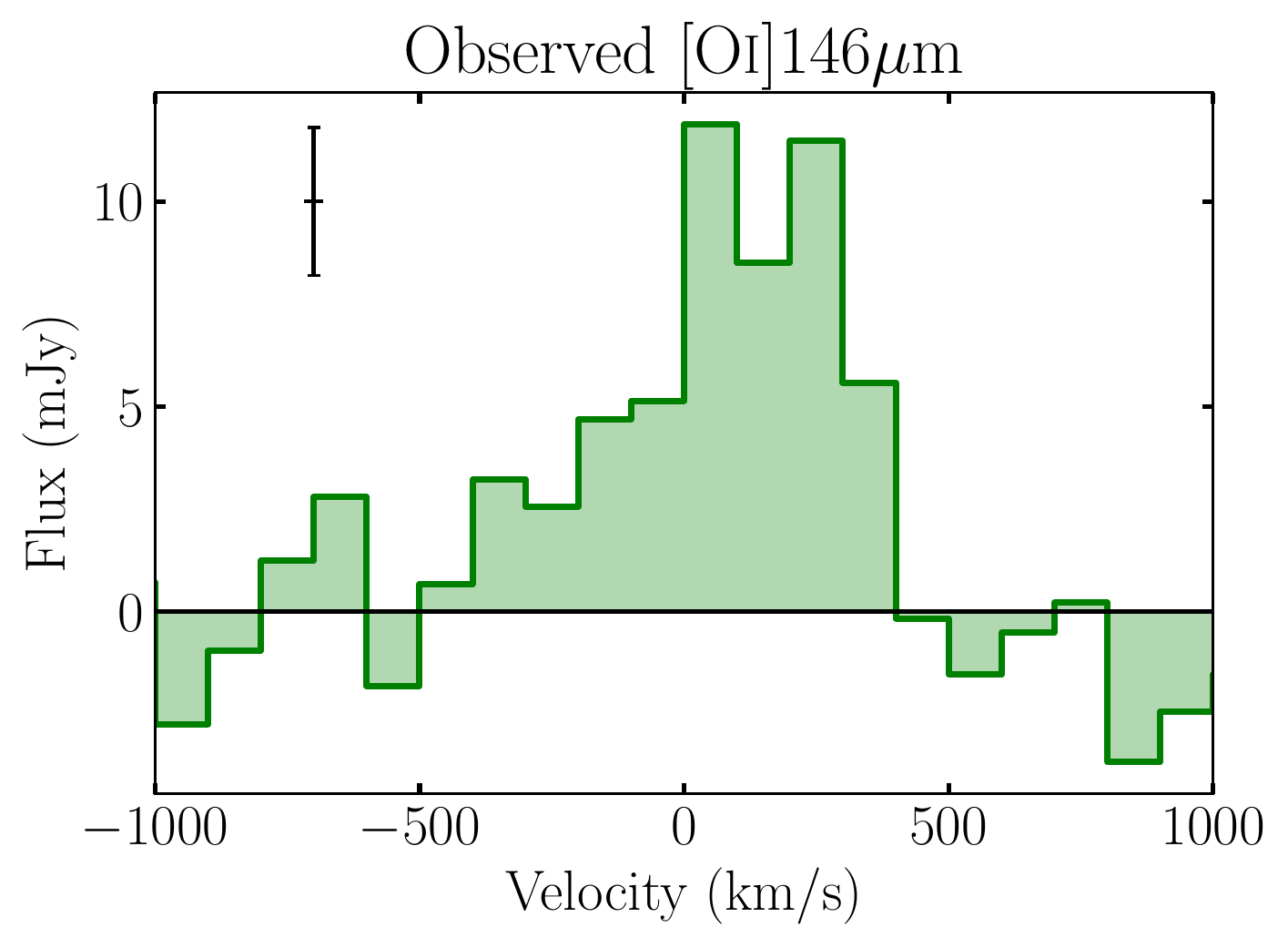}
\includegraphics[width=0.32\linewidth]{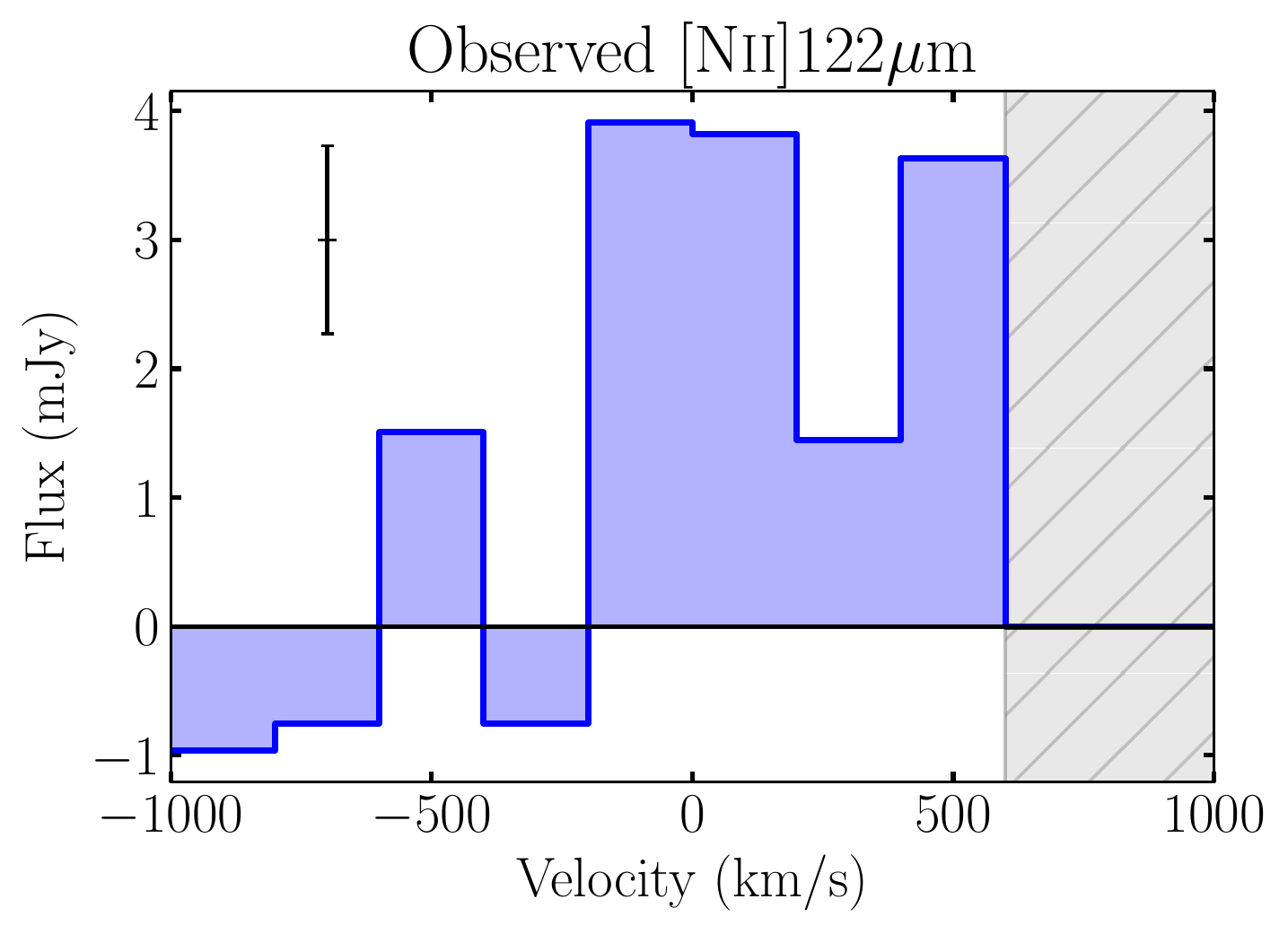}
\includegraphics[width=0.32\textwidth]{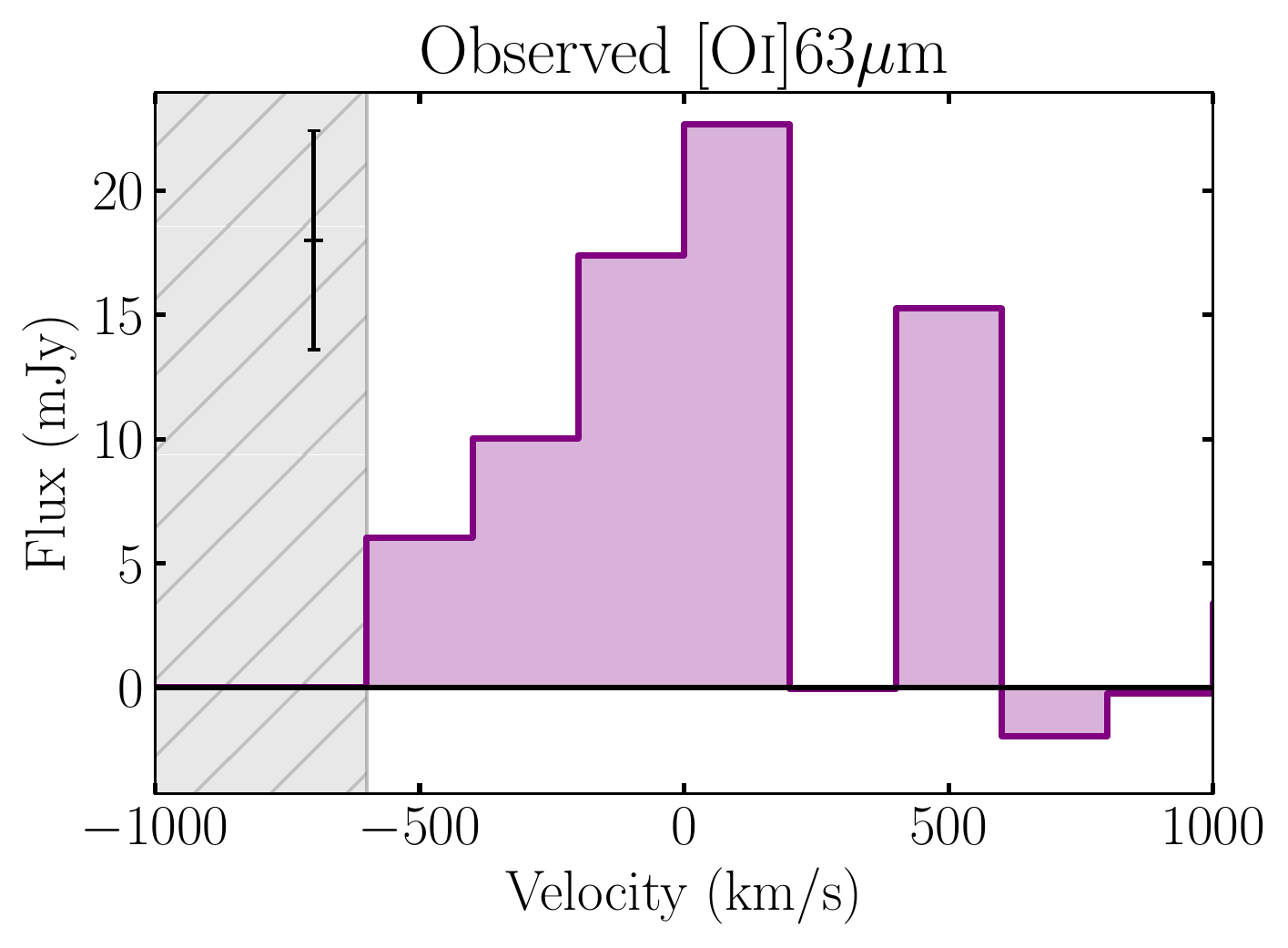}
\includegraphics[width=0.32\textwidth]{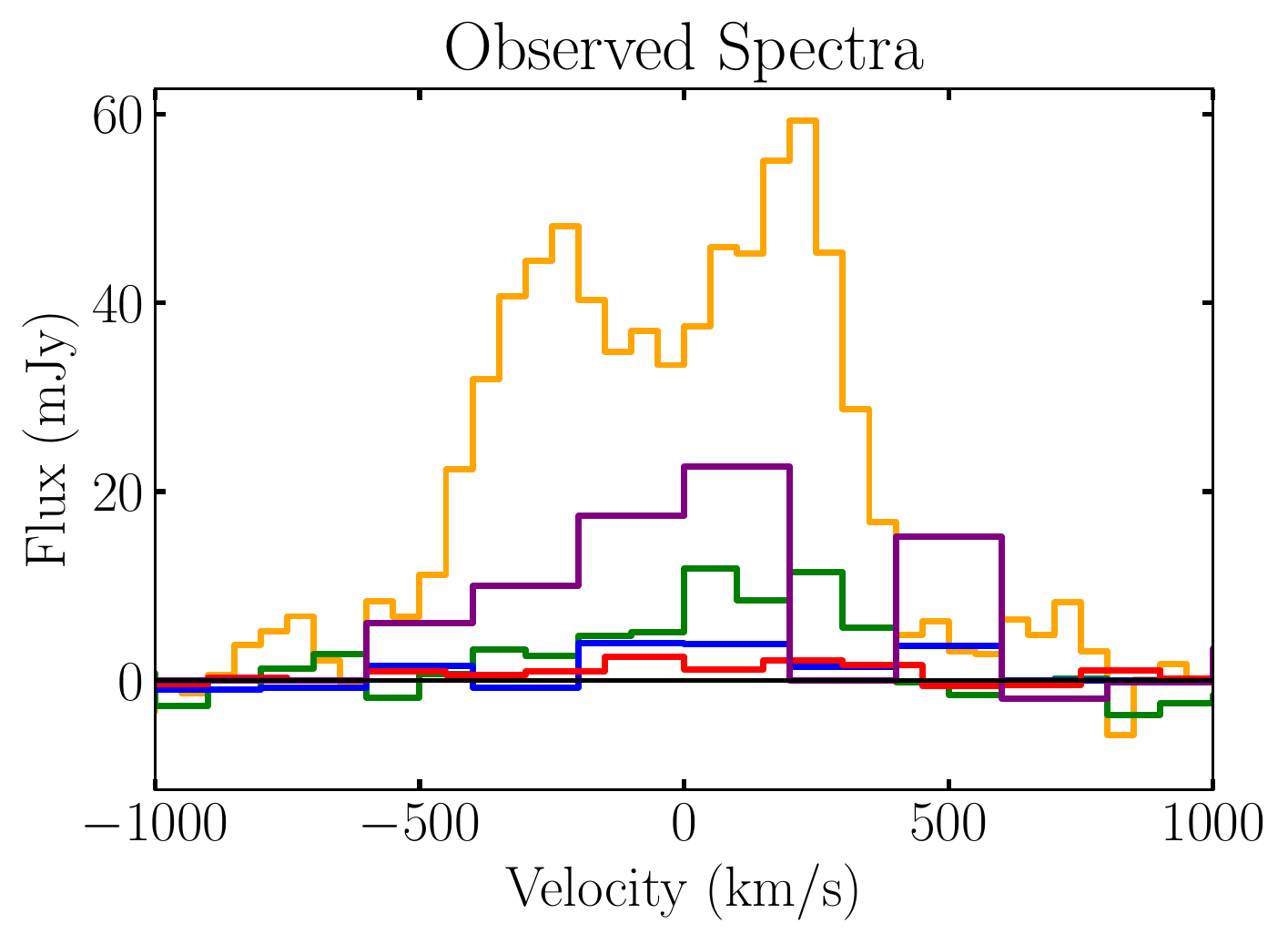}
\caption{Observed spectra of the targeted lines. Top row (left to right): \nii 205\micron, \cii 158\micron, \oi 146\micron.  Bottom row (left and center): \nii 122\micron, \oi 63\micron.  The rightmost bottom panel shows all five spectra overlaid on the same vertical scale.  The grey regions in the \nii 122\micron\ and \oi 63\micron\ spectra represent velocities between spectral windows.  The spectra were obtained using observed and model visibilities, as described in the appendix of \cite{litke2019}.
Typical uncertainties are plotted in the upper left corners of the \nii 205\micron, \cii 158\micron, \oi 145\micron, \nii 122\micron, and \oi 63\micron\ spectra.
\label{fig:obsspec}}
\end{center}
\end{figure*}

To evaluate the overall significance of our line detections, we construct source-integrated spectra from the visibility data.
Because SPT0346-52 is gravitationally lensed, we use the spatial structure of the continuum emission to provide a spatial template for the line emission (see the Appendix of \cite{litke2019}). Visibilities of the line emission are weighted by a gravitational lensing model of the continuum emission to emphasize the visibilities that best sample the source structure, yielding a channelized flux density ($F_{\nu}$) determined by
\begin{equation}
\label{eq:spec}
F_v = \frac{\sum_i \frac{\widetilde{v}_{v,i}}{\widetilde{m}_i} \vert \widetilde{m}_i\vert ^2}{\sum_i \vert \widetilde{m}_i\vert ^2}.
\end{equation}
Here, 
$\widetilde{v}_{v,i}$ is the complex line data visibility and
$\widetilde{m}_i$ is the complex model visibility for that data set.
The model visibilities are obtained from our lensing reconstructions, described in Section \ref{sec:lens}.
The observed spectrum for each line is shown in Figure \ref{fig:obsspec}.
To obtain the uncertainties, visibilities were randomly drawn from the distribution of visibilities for each channel and each line 500 times.
The random spectra were then calculated using Equation \ref{eq:spec}.  The uncertainties were then determined by taking the standard deviation of the 500 random noise trials.

\nii 122\micron\ is not significantly detected in our observations.  \nii 122\micron\ is redshifted to 369.5GHz, where the atmospheric transmission declines due to a strong atmospheric O$_2$ line at 368.5GHz.
\oi 63\micron\ is also not significantly detected in our observations. This line is redshifted to 712.9GHz, the high-frequency end of ALMA Band 9, where a strong atmospheric O$_2$ line at 715.4 GHz and the 752 GHz water line that separates the 650 and 850 GHz atmospheric windows (Bands 9 and 10) combine to produce a sharp decline in atmospheric transmission toward higher frequency (bluer velocity). 
To obtain the upper limits for \nii 122\micron\ and \oi 63\micron\ listed in Table \ref{table:mcmc}, the $1\sigma$ uncertainty on a single 600 km/s channel was calculated using the method described above.
This value was multiplied by 3 to obtain the $3\sigma$ upper limit.  It was then divided by $\mu = 5.6$ \citep{spilker2016} to correct for magnification from gravitational lensing.  The non-detections of \nii 122\micron\ and \oi 63\micron\ are discussed in Sections \ref{sec:nii} and \ref{sec:oi63}, respectively.

\section{Lens Modeling}
\label{sec:lens}

Gravitational lensing is a powerful tool for studying galaxies at high redshift.
Because lensing spreads the source emission over a larger solid angle on the sky while preserving the surface brightness, resolving detail in the lensed galaxy can be done with a more compact array configuration than is possible in unlensed sources, and more compact arrays have better surface brightness sensitivity.
However, the image distortion introduced by the gravitational lensing makes it make it difficult to study the spatially-resolved physical structure of the galaxy in a straight-forward manner.

In order to study the source-plane structure of SPT0346-52, we turn to lensing reconstruction models.
We use a pixelated, interferometric lensing reconstruction code, \textsc{ripples} \citep{hezaveh2016}.  Additional information in the general framework for pixelated lens modeling is described by \cite{warren2003} and \cite{suyu2006}.
\textsc{ripples} uses a Markov Chain Monte Carlo (MCMC) method to model the mass distribution of the foreground galaxy as well as the background source emission.
It also takes into account observational effects from the primary beam.
In addition, a regularization factor is introduced that minimizes large gradients between adjacent pixels, which prevents overfitting of the data.

As mentioned above, \textsc{ripples} models the complex visibilities observed by ALMA directly, rather than modeling \textsc{clean} images.
By modeling the complex visibilities, we use all of the data available from the ALMA observations.
Because \textsc{ripples} is a pixelated code, we do not assume a source-plane structure, 
and can model more complex structures.

\begin{deluxetable*}{rccccc}
\centering
\tablecaption{SPT0346-52 Continuum Lens Modeling Results}
\tablenum{2}
\label{table:mcmc}
\tablehead{\colhead{Parameter} & \colhead{205\micron} & \colhead{158\micron} & \colhead{146\micron} & \colhead{122\micron} & \colhead{63\micron}}
\startdata
\textbf{Lens Parameters} & & & & & \\
$\log$Mass[\msol] & $11.46\pm 0.01$ & $11.46\pm 0.01$ & $11.46\pm 0.02$ & $11.50\pm 0.01$ & $11.47\pm 0.02$ \\
Ellipticity x-Component, $e_x$ & $-0.17\pm 0.01$ & $-0.17\pm 0.01$ & $-0.18\pm 0.02$ & $-0.16\pm 0.01$ & $-0.16\pm 0.01$ \\
Ellipticity y-Component, $e_y$ & $+0.43\pm 0.03$ & $+0.41\pm 0.04$ & $+0.39\pm 0.07$ & $+0.22\pm 0.01$ & $+0.39\pm 0.09$ \\
Ellipticity, $e^{\mathrm a, c}$ & 0.45 & 0.45 & 0.42 & 0.25 & 0.45 \\
Position Angle, $\phi_e$ (E of N)$^{\mathrm b, c}$ & 69$^\circ$ & 68$^\circ$ & 65$^\circ$& 55$^\circ$ & 68$^\circ$ \\
Shear x-Component, $\gamma_x$ & $+0.06\pm 0.01$ & $+0.06\pm 0.01$ & $+0.09\pm 0.02$ & $+0.15\pm 0.01$ & $+0.07\pm 0.03$ \\
Shear y-Component, $\gamma_y$ & $-0.10\pm 0.01$ & $-0.11\pm 0.01$ & $-0.10\pm 0.01$ & $-0.08\pm 0.01$ & $-0.10\pm 0.01$ \\
Shear Amplitude, $\gamma^{\mathrm a, c}$ & 0.12 & 0.12 & 0.13 & 0.17 & 0.12 \\
Shear Position Angle, $\phi_{\gamma}$ (E of N)$^{\mathrm b, c}$ & 120$^\circ$ & 120$^\circ$ & 127$^\circ$ & 115$^\circ$ & 119$^\circ$ \\
Lens x Position, $x$ & $0\farcs04 \pm 0\farcs01$ & $0\farcs08 \pm 0\farcs01$ & $0\farcs00 \pm 0\farcs01$ & $0\farcs01 \pm 0\farcs01$ & $0\farcs05 \pm 0\farcs01$ \\
Lens y Position, $y$ & $0\farcs38 \pm 0\farcs01$ & $0\farcs32 \pm 0\farcs01$ & $0\farcs40 \pm 0\farcs01$ & $0\farcs39 \pm 0\farcs01$ & $0\farcs35 \pm 0\farcs01$ \\
\hline
\textbf{Source-Plane Fluxes} &  &  &  &  &  \\ 
Continuum Flux (mJy) & $6.8\pm 1.1$ & $11.9\pm 1.2$ & $13.2\pm 1.3$ & $24.9\pm 2.5$ & $28.1\pm 2.8$ \\
Line Luminosity ($10^8$ \lsol) & $1.2 \pm 0.2$ & $34 \pm 5$ & $3.2 \pm 0.5$ & $< 3.5^{\mathrm d}$ & $< 53^{\mathrm d}$ \\
\enddata
\tablenotetext{a}{$\alpha = \sqrt{\alpha_x^2+\alpha_y^2}$, where $\alpha = e$ or $\alpha = \gamma$}
\tablenotetext{b}{$\phi_{\alpha} = \arctan{(-\alpha_y/\alpha_x)}$, where $\alpha= e$ or $\alpha = \gamma$}
\tablenotetext{c}{Derived from best-fit parameters}
\tablenotetext{d}{$3\sigma$ upper limit from observations, corrected for lensing}
\end{deluxetable*}

\begin{figure*}
\begin{center}
\includegraphics[width=0.9\textwidth]{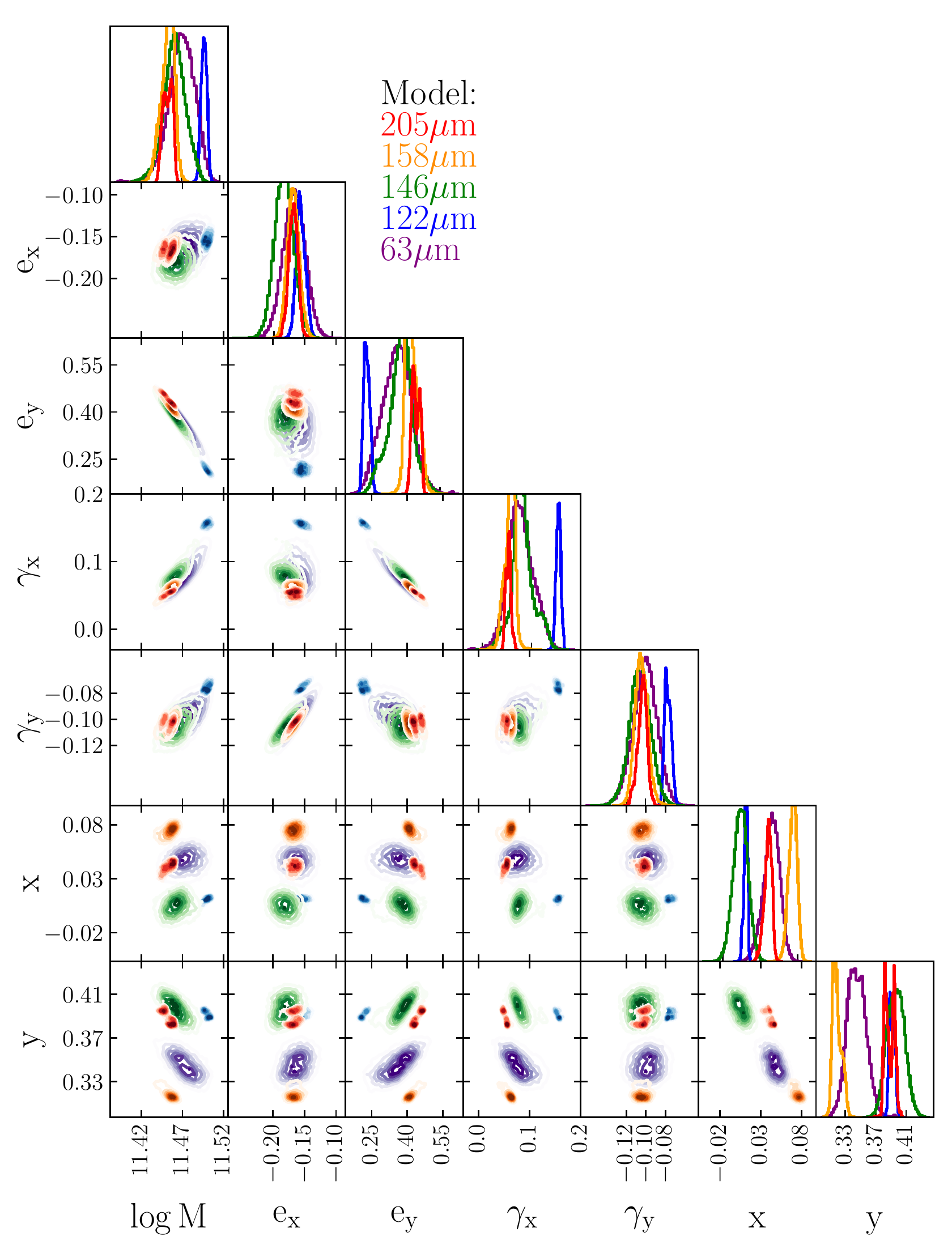}
\caption{Triangle plot with the model lens parameters for SPT0346-52 computed for different continuum wavelengths.  Red:  205\micron \ continuum model.  Orange:  158\micron \ continuum model.  Green:  146\micron \ continuum model.  Blue:  122\micron \ continuum model.  Purple:  63\micron \ continuum model.  $M$ is the lens mass enclosed within 10 kpc and is measured in \msol.  $e_x$ and $e_y$ are the two components of the lens galaxy's ellipticity, while  $\gamma_x$ and $\gamma_y$ are the two components of the shear.  $x$ and $y$ are the offset of the lens center from the phase center in arcseconds.}
\label{fig:triangle}
\end{center}
\end{figure*}

We model the ALMA observations using the same procedure as \cite{litke2019}.
The mass distribution of the foreground lensing galaxy is modeled as a singular isothermal ellipsoid (SIE) at $z=0.9$ with an external shear component.  Previous lens modeling of SPT0346-52 by \cite{hezaveh2013}, \cite{spilker2015}, and \cite{litke2019} were used to obtain the initial parameters.
The 205\micron, 158\micron, 146\micron, 122\micron, and 63\micron\ rest-frame continuum data were all fit independently.  The best-fit lens model derived from each continuum data set was then applied to the corresponding line data corresponding to that continuum.  For example, the 158\micron \ model was applied to the \cii 158\micron\ line data, while the 205\micron \ model was applied to the \nii 205\micron\ line data.
The lines were integrated from -300 km/s to +300 km/s for the reconstructions.
Table \ref{table:mcmc} gives the best-fit lensing parameters for all five models, as well as the source-plane continuum fluxes and line luminosities.
The parameter covariance plot from the MCMC for all five continuum sets is shown in Figure \ref{fig:triangle}.

\subsection{Comparing Models}
\label{sec:modcomp}

As seen in Figure \ref{fig:triangle}, the five continuum models do not have identical lens parameters, though they are very similar in most parameters.
\footnote{The exception lies with the lens x and y positions.  These positions are measured relative to the phase-centers of observations.  The variation in the modeled positions of the lens from different datasets are consistent with the variation seen in the position of the astrometric test sources.}
In order to explore the effect of the differing models on the source-plane reconstructions, we applied each model to each of the other continuum data sets.
We then reconstructed the source-plane continuum emission.
The derived continuum fluxes were consistent, independent of the model used.
The differences between the best-fit models are within the errors of the MCMC fit and most likely result from degeneracies between the ellipticity and shear parameters.
These differences do not affect our results in the source-plane.

To find the uncertainty on the flux in each pixel, we created 500 sets of random visibilities from the distribution of uncertainties in the visibilities.  We then reconstructed these 500 random noise data sets and took the standard deviation in each pixel.  The total error used is the standard deviation of the random noise reconstructions added in quadrature with 10\% of the flux.

In order to determine the effective resolution of the reconstructed maps, we follow the method used by \cite{litke2019}.
We define the effective resolution as the inferred source-plane size when reconstructing a lensed point source at that source-plane position.
  This effective resolution will vary depending on the position of the source relative to the caustic in the source-plane, as well as the signal-to-noise of the input data.
For each set of visibilities, we create a point sources with the flux and position of the emission at that wavelength.  We then apply the corresponding lens model in Table \ref{table:mcmc} to these point sources to create a lensed set of visibilities.  Next, we make source-plane reconstructions of these lensed point sources using \textsc{ripples}.
Finally, we fit a 2D-Gaussian to the reconstructed image to find the effective resolution.

\begin{figure*}
\begin{center}
\includegraphics[width=\textwidth]{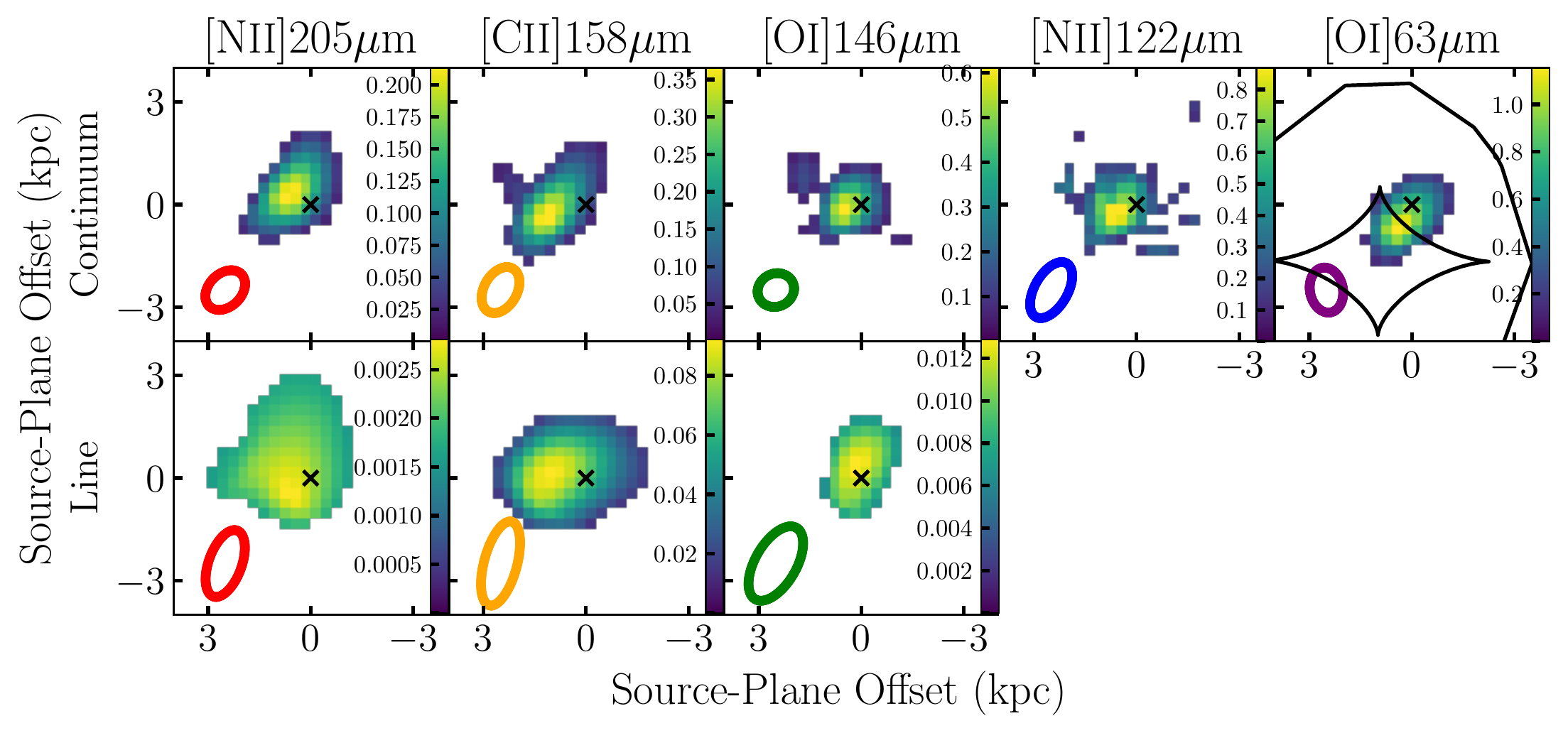}
\caption{Reconstructed source-plane emission.  From left to right: 205\micron, 158\micron, 146\micron, 122\micron, and 63\micron.  Top:  continuum maps of the source-plane. Colorbar units are mJy.
Bottom Row:  Moment 0 maps (velocity-integrated line flux).  Colorbar units are mJy km/s.  The ellipses represent the typical resolution where there is emission in the source-plane.
The black x's mark the center of the source plane.  A representative caustic is shown as black lines in the upper right corner (the 63\micron\ continuum map). 
 All source-plane images are 8kpc per side.
\label{fig:mommaps}}
\end{center}
\end{figure*}

It is simpler to compare the different line and continuum reconstructions if they have comparable resolutions.
Therefore, we tapered the visibilities in each data set to $300k\lambda$ before performing the reconstructions.
The resolutions are shown as colored ellipses in the lower left corner of the continuum and Moment 0 maps in Figure \ref{fig:mommaps}.
The effective resolution is $\sim 0\farcs08 \times 0\farcs 15$, which is gives us $\sim 700\ \rm{pc}$ resolution.
The models, residuals, and error maps are shown in Figure \ref{fig:residual} in Appendix \ref{app:lens}.

\subsection{Lens Modeling Results}
\label{sec:lensresults}

Figure \ref{fig:mommaps} shows the reconstructed continuum maps and the Moment 0 (integrated flux)maps of the line reconstructions.
The continuum maps mostly show similar morphologies, but with different fluxes.  As expected, the 63\micron\ continuum emission is the brightest and the 205\micron\ emission is the weakest.
We see differing morphologies in the line emission.
The \cii 158\micron\ and \nii 205\micron\  lines show their brighest emission offset from the \oi 146\micron\ emission.
Because the Earth's atmosphere limits access to these lines except at the highest redshifts, the sources with spatially resolved maps of FIR fine structure lines are mostly very nearby galaxies. 
\cite{parkin2013} and \cite{herreracamus2018a} found in M51 and NGC1068, respectively, that the \cii 158\micron\ and \nii\ lines had similar morphologies.
A similar offset between \oi 146\micron\ and \cii 158\micron\ emission was seen by \cite{herreracamus2018a} in NGC1068.
  In their maps, the brightest \oi 146\micron\ emission was associated with the central AGN in NGC1068, while the \cii 158\micron\  emission was associated with the peak in CO emission.  
\cite{parkin2013} also found that the \oi 63\micron\ and \oi 146\micron\ emission peaked in the center of M51, most likely associated with the central AGN, offset from where \cii 158\micron\ and \nii 122\micron\ emission was strongest.  
In the SMC, \oi 63\micron\ emission was associated with H$\alpha$ emission and therefore recent massive star formation \citep{jameson2018}.
As seen in these other systems, the \oi 146\micron\ and \oi 63\micron\  emission has been associated with regions where one would expect dust heating, whether the dust is being heated by star formation or a central AGN.
For SPT0346-52, the \oi 146\micron\ emission is concentrated closer to where the dust continuum is strongest. 
As any potential AGN contribution to the IR emission in SPT0346-52 is negligible \citep{ma2016}, the \oi 146\micron\ emission appears to coincide with the most intense star formation.

\cite{litke2019} determined that the intense star-formation in SPT0346-52 was driven by a major merger of two components.  These components are centered at -310 and +160 km/s and have similar widths ($\sim 300$ km/s).  The center of these two components lie near -75 km/s.
The merger status was determined using position-velocity diagrams of the \cii 158\micron\  emission, which is the highest signal-to-noise line of those explored here.
\cite{litke2019} used higher resolution reconstructions than are shown in this work, where we decrease the resolution of the \cii 158\micron\  reconstruction to match that of the lowest resolution lines.
More recently, \cite{jones2019} claimed SPT0346-52 is better described as a disk galaxy with a molecular outflow, based on visual inspection of the image-plane structure.  However, the candidate H$_2$O outflow \cite{jones2019} detected is kinematically similar to the blue-shifted component found by \cite{litke2019}.  The mass loading factor of the possible outflow in SPT0346-52 is well below unity, unlike other outflows with mass loading factors near or greater than unity.
\cite{spilker2020} do not see broad wings in \cii 158\micron\  that would be indicative of outflows in any DSFGs with confirmed molecular outflows traced by blue-shifted OH absorption.
Thus, SPT0346-52 is unlikely to host outflow activity.
The disk structure described by \cite{jones2019} may also have resulted from a recent major merger \citep[e.g.,][]{hopkins2009}.

\section{CLOUDY Modeling}
\label{sec:cloudy}

In order to understand the physical conditions that explain our observations, we turn to the photoionization code \textsc{Cloudy} \citep[version 17.01;][]{ferland2017}.
\textsc{Cloudy} simulates the microphysics within a cloud of gas and dust that is heated by a central source.  It predicts the physical conditions throughout the cloud, including temperatures, densities, and metallicities, while also computing a predicted observed spectrum.

\subsection{CLOUDY Parameters}
\label{sec:cloudyparam}

We model our system as an open, or slab-like geometry.  We adopt an inner radius (the distance between the central heating source and the inner face of the cloud) of 100 pc.
This distance is chosen to be smaller than the size of an individual pixel.

We use the ISM gas-phase elemental abundances and grain size distributions included with \textsc{Cloudy}, which represents the average warm and cold phase abundances of the ISM in the Milky Way \citep{cowie1986,savage1996}.
We also include polycyclic aromatic hydrocarbons (PAHs) in the simulation.
Small grains such as PAHs are an important contributor of grain heating mechanisms and FUV radiative transfer effects \citep{hollenbach1999}.
For the equation of state of the ISM, we assume constant pressure.
The components balanced to achieve constant pressure gas are
\begin{equation}
P_{tot} = P_{gas} + P_{turb} + P_{lines} + \Delta P_{rad},
\end{equation}
where $P_{gas}$ is thermal gas pressure,
$P_{turb}$ is the turbulent pressure,
$P_{lines}$ is radiation pressure due to trapped emission lines,
 and $\Delta P_{rad}$ is pressure from the attenuation of the incident radiation field.
In order to simplify the model, we do not include a magnetic field component.
Following \cite{cormier2019}, we assume a constant microturbulent velocity of 1.5 $\rm km/s$.  This is consistent with the microturbulent velocities of individual PDRs \citep{kaufman1999,tielens1985}.

The gas cloud is heated by a single-burst stellar population.
The starburst spectral energy distribution (SED) was compiled by \cite{byler2017} for use in \textsc{Cloudy} using the Flexible Stellar Population Synthesis code \citep{conroy2009,conroy2010}.
We use the ionizing spectrum from \cite{byler2017} produced by the MESA Isochrones and Stellar Tracks \citep[MIST;][]{choi2016,dotter2016}.
The MIST stellar evolution tracks differ from other models in that they include stellar rotation, which results in harder ionizing spectra and higher luminosities \citep{byler2017}.
\cite{byler2017} compared nebular emission ionized by MIST models to Padova \citep[for low-mass stars;][]{bertelli1994,girardi2000,marigo2008} and Geneva \citep[for high-mass stars;][]{schaller1992,meynet2000} evolutionary tracks \citep{levesque2010}, and found that the MIST models can match observed line ratios better as the starburst ages past a few Myr.
We fix the stellar metallicity to $\log{Z/Z_{\odot}}=0$ and the age to the stellar age calculated by \cite{ma2015} using SED fitting, $\sim 30$ Myr.
Allowing the stellar metallicity to scale with the gas metallicity did not change our results.
The input starburst SED sets the shape of of the ionizing spectrum in \textsc{cloudy}.  The intensity of the ionizing radiation is determined by the ionization parameter, described below.

We also include the cosmic microwave background (CMB) at $z=5.7$ and cosmic rays to contribute to the gas heating.
The CMB spectrum was then subtracted from the modeled continuum SED before comparing to the observations, which are measurements of excess above the CMB spectrum.

We create a grid of models varying the ionization parameter ($U$), Hydrogen density 
($n_H$) at the face of the cloud, gas metallicity ($Z$), and the age of the starburst.
The ionization parameter is defined as the ratio of hydrogen-ionizing photons to the total hydrogen density, or more specifically
\begin{equation}
U \equiv \frac{Q(H)}{4 \pi r_0^2 n_H c},
\end{equation}
where $r_0$ is distance between center of starburst and inner surface of cloud (100 pc),
$n_H$ is the total hydrogen density,
$Q(H)$ is the number of hydrogen-ionizing photons, and $c$ is the speed of light.
We vary $\log{U}$ from -4.5 to -0.5 in steps of 0.25.
The total hydrogen density includes molecular, atomic, and ionized hydrogen components.
$n_H$ is defined at the inner face of the cloud for the grid values $0.5 \le \log{n_H} \le 3.5$ in steps of 0.25.
We vary the gas metallicity from $-2.0 \le \log{Z/Z_{\odot}} \le 2.0$ in steps of 0.25. 
We stop the \textsc{cloudy} simulation at visual extiction $A_V = 100$, following \cite{abel2009}.

To find the best-fit model, we compare the \textsc{cloudy} outputs to a combination of continuum ratios ($\log$ 63\micron/122\micron, $\log$ 63\micron/146\micron, $\log$ 63\micron/158\micron, $\log$ 63\micron/205\micron), line ratios ($\log$\cii 158\micron/\nii 205\micron, $\log$\cii 158\micron/\oi 146\micron), and a line-to-continuum ratio ($\log$\oi 146\micron/146\micron) to constrain the relative contributions of gas and dust.  
The best-fit model is chosen to be the model with the lowest reduced chi-squared value, $\chi^2_r$.
Continuum values are in mJy and line values are in \lsol.  
\oi 146\micron/146\micron\ has units $10^7$ \lsol/mJy.  The factor of $10^7$ is used so the line-to-continuum ratio is the same order of magnitude as the continuum ratios and line ratios.
Table \ref{table:ratios} in the Appendix lists the values used to compare to the \textsc{cloudy} models.
We consider both spatially-resolved and galaxy-integrated emission models.

We tested whether the inclusion of the continuum ratios affected our results.  When the continuum ratios were not included, the line ratios were still well-fit, but the continuum ratios from \textsc{cloudy} did not match our observations.  The ionization parameter was the variable most sensitive to the inclusion of the continuum ratios.
This is consistent with previous \textsc{cloudy} modeling, where \cite{abel2009} found the 60\micron/100\micron\ continuum ratio was strongly dependent on the ionization parameter.

\subsection{Cloudy Modeling Results}
\label{sec:cloudyresults}

\begin{figure}
\begin{center}
\includegraphics[width=0.45\textwidth]{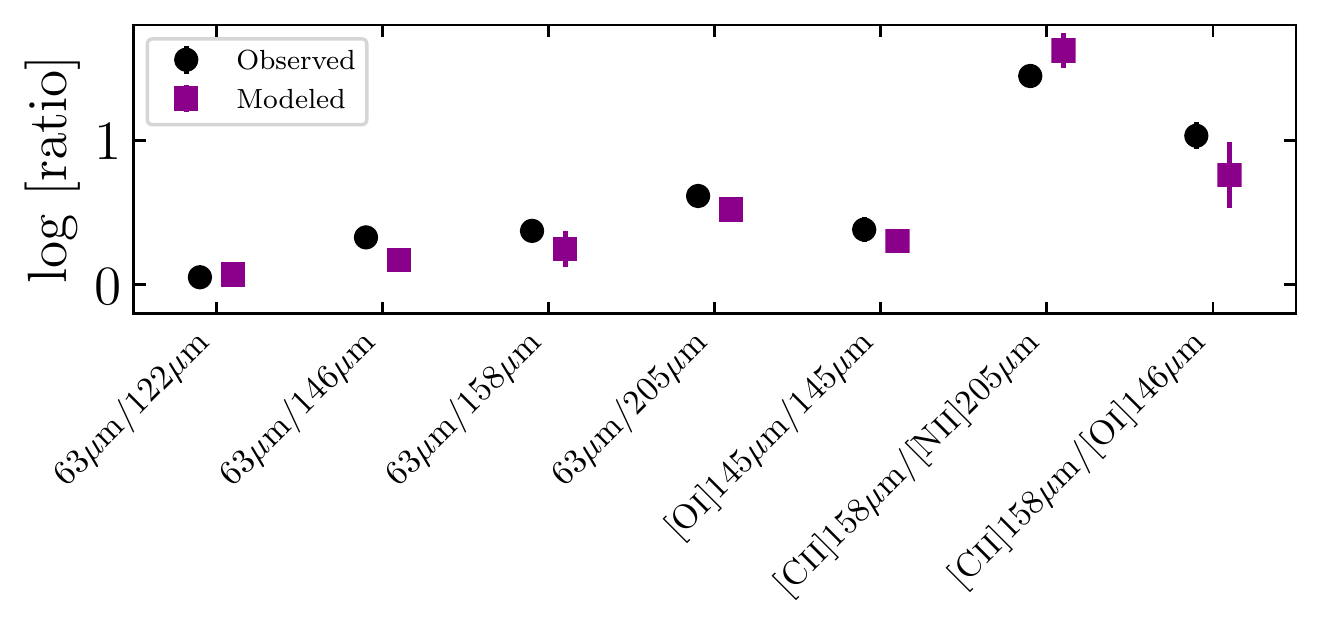}
\caption{Comparison of observed continuum and line ratios to \textsc{cloudy} best-fit galaxy-integrated ratios.  
From left to right:  $\log$ 63\micron/122\micron, $\log$ 63\micron/146\micron, $\log$ 63\micron/158\micron, $\log$ 63\micron/205\micron, $\log$ \oi 146\micron/146\micron, $\log$\cii 158\micron/\nii 205\micron, $\log$\cii 158\micron/\oi 146\micron.
Black circles represent the observed ratios and associated uncertainties.  Purple squares represent the best-fit ratios from the \textsc{cloudy} modeling and associated uncertainties.  
\label{fig:cloudyratiosg}}
\end{center}
\end{figure}

Figure \ref{fig:cloudyratiosg} shows the observed continuum and line ratios with their associated errors, as well as the ratios calculated for the best-fit \textsc{cloudy} model for SPT0346-52.
Table \ref{table:cloudy} lists the best-fit starburst age, $U$, $n_H$, and $Z$ values, as well as $\chi^2_r$, for the best-fit models for the global fits shown in Figure \ref{fig:cloudyratiosg}.
Overall, the best-fit models agree with the observed global line and continuum ratios in SPT0346-52.

\begin{deluxetable}{rc}
\centering
\tablecaption{Best-Fit Global \textsc{cloudy} Models$^{\mathrm a}$}
\tablenum{3}
\label{table:cloudy}
\tablehead{
\colhead{Parameter} & \colhead{Value}}
\startdata
$\log{U}$ & $-2.75^{+1.3}_{-0.1}$\\
$\log{n_H [cm^{-3}]}$ & $1.75^{+0.1}_{-1.1}$\\
$\log{Z/Z_{\odot}}$ & $0.75^{+0.5}_{-0.1}$\\
$\chi^2_r$ & 6.6\\
\enddata
\tablenotetext{a}{The likelihood distribution is calculated by summing $e^{-\chi_r^2}$.  The best-fit value is the peak of the likelihood distribution, and the uncertainties are  where the likelihood distribution is within 1$\sigma$ of the peak.}
\end{deluxetable}

\begin{figure}
\begin{center}
\includegraphics[width=0.45\textwidth]{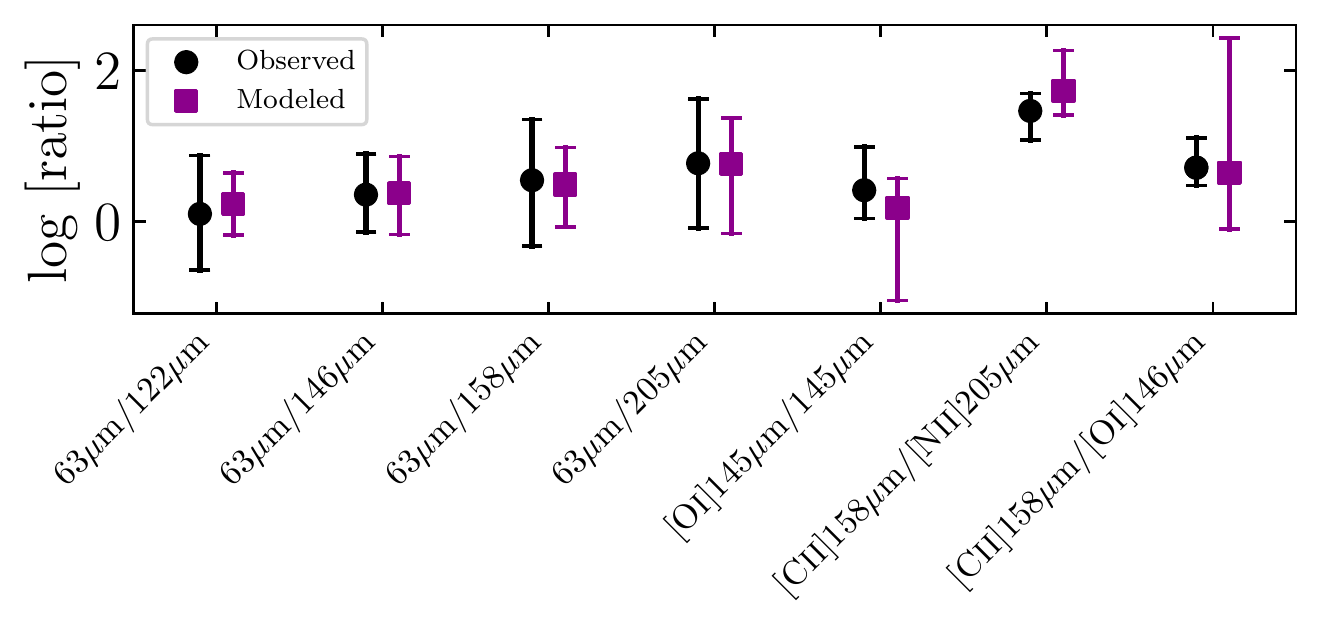}
\caption{Comparison of observed continuum and line ratios to \textsc{cloudy} best-fit pixelated ratios.  From left to right:  $\log$ 63\micron/122\micron, $\log $63\micron/146\micron, $\log $63\micron/158\micron, $\log$ 63\micron/205\micron,  $\log$\oi 146\micron/146\micron, $\log$\cii 158\micron/\nii 205\micron, $\log$\cii 158\micron/\oi 146\micron.  Black circles represent the means of the observed pixel ratios, and the errorbars represent the maximum and minimum observed ratios.  Purple squares represent the means of the best-fit pixel ratios  from the \textsc{cloudy} modeling, with the maximum and minimum best-fit ratios.
\label{fig:cloudyratiosp}}
\end{center}
\end{figure}

\begin{figure}
\begin{center}
\includegraphics[width=0.4\textwidth]{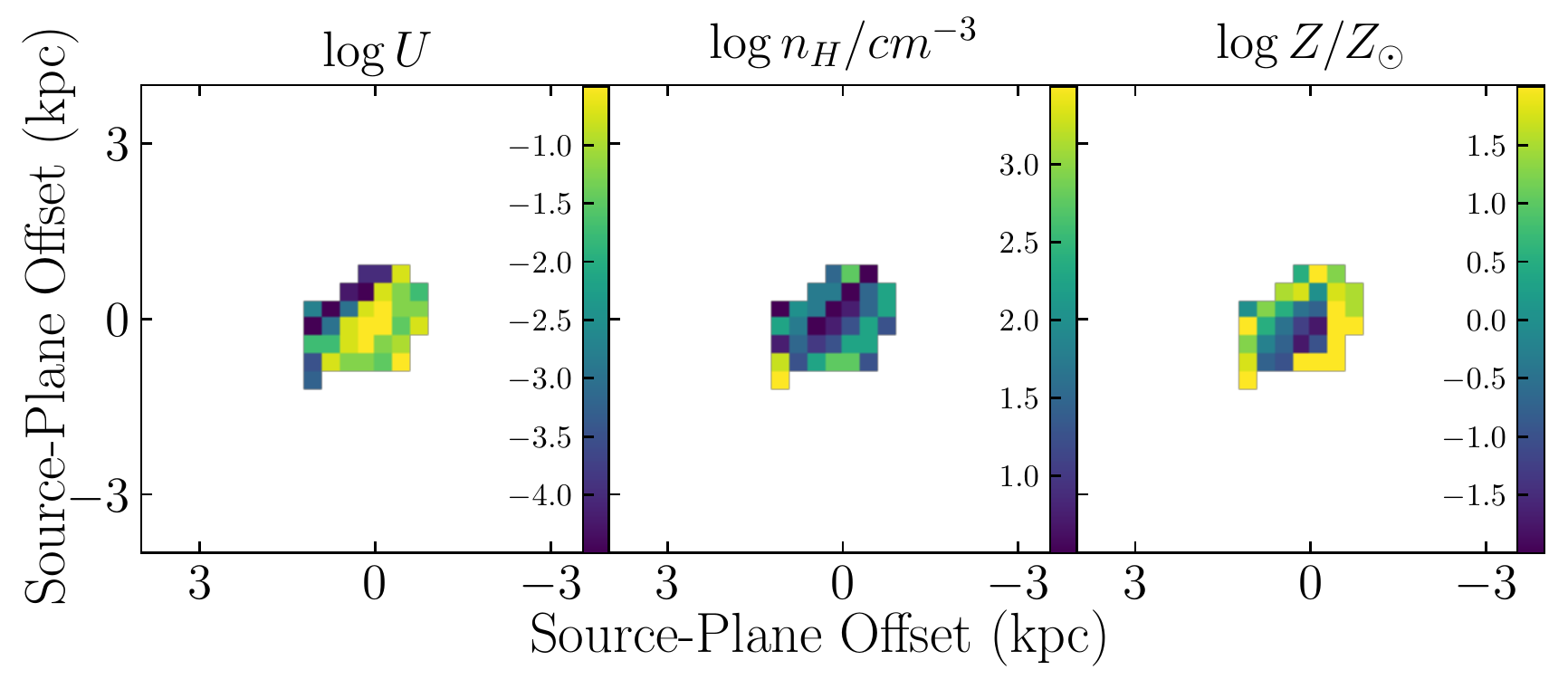}
\caption{Best-fit spatially-resolved \textsc{cloudy} parameters.  Left:  $\log U$.  Middle:  $\log n_H/cm^{-3}$.  Right:  $\log{Z/Z_{\odot}}$.
\label{fig:cloudyparams}}
\end{center}
\end{figure}

It is a crude approximation to model an entire galaxy as a single PDR. Indeed, \cite{katz2019} found in their simulations of emission lines in high-z galaxies that there can be very large ranges in metallicty and ionization parameter across a single galaxy.
 In the case of SPT0346-52, we have spatially-resolved information about the line and continuum emission and can explore the distribution of properties across the galaxy by fitting our models on a pixel-by-pixel basis.
We use the same technique as was used for the galaxy-integrated fits for each pixel.  Figure \ref{fig:cloudyratiosp} shows the range of values observed and in the best-fit models, as well as the mean values, for each set of continuum and line ratios, while Figure \ref{fig:cloudyparams} shows the best-fit $U$, $n_H$, and $Z$ values in each pixel.
 For maps of the observed continuum and line ratios, as well as the best-fit model ratios, see Figure \ref{fig:cloudymaps} in the Appendix.

As with the fits to the global line and continuum ratios, the range of ratios in the pixellated best-fit models mostly agree with the range of ratios observed.
 The modeled \cii 158\micron/\nii 205\micron\  ratios have higher maximum values than the observed ratios, though their average values are more similar.  We also see a higher average \cii 158\micron/\oi 146\micron\ model ratio than observed, as we did with the galaxy-integrated fit.  This may result from the \textsc{cloudy} model not probing far enough from the ionizing source to fully recover line emission from the neutral gas component of the ISM.

The best-fit ionization parameters are around $\log{U} = -2.75$ and
 is relatively uniform.
These values are comparable to those found for the DSFG SPT0418-47 \citep[using \nii 205, \cii 158, \oi 146, \nii 122, and \oiii 88;][]{debreuck2019}.
A popular technique is to report the intensity of the FUV, $G$, relative to the interstellar radiation field, $G_0$, instead of the ionization parameter.  To compare our $U$ values to models using $G$, we turn to Figure 1 of \cite{abel2009}; for a starburst SED, $\log{U} \approx \log{G} - 6$ and $\log{U} = -3.25$ corresponds to $G = 10^{3.25} G_0$.
\cite{rybak2019} and \cite{rybak2020} found $\log{G/G_0} \approx 4$ for $z\sim 3$ DSFGs and the $z=6$ DSFG G09.83808, respectively, while \cite{novak2019} found $\log{G/G_0} > 3$ for the $z=7.5$ quasar host galaxy J1342+0928.
SPT0346-52 has a lower ionization parameter and FUV field strength compared to other high-z DSFGs.

We find densities around $\log{n_H} \sim 2$ throughout this system.
This is lower than the inferred densities of the DSFGs SPT0418-47 at $z=4.225$ \citep[$\log{n} \sim 4.3$;][]{debreuck2019} and G09.83808 at $z=6.027$ \citep[$\log{n} \sim 4$;][]{rybak2020}.

From the \textsc{cloudy} modeling, SPT0346-52 appears to have a supersolar metallicity ($\log Z/Z_{\odot} = 0.75$).
This is higher than other high-z sources that have been found to have metallicities near solar ($\log{Z/Z_{\odot}} \sim 0.1$ for quasar-host galaxy J1342+0928 at $z=7.54$ and $-0.5 < \log{Z/Z_{\odot}} < 0.1$ for DSFG SPT0418-47) using \oiii 88/\nii 122 \citep{novak2019,debreuck2019}.
However, the \oiii 88/\nii 122 metallicity diagnostic is highly dependent on the ionization parameter \citep{pereirasantaella2017}.
Using the mass-metallicity relation of elliptical galaxies, \cite{tan2014} found the metallicity of DSFGs in the protocluster GN20 ($z=4.05$) to be $\log{Z/Z_{\odot}} \sim 0.5 \pm 0.2$.
The gas to dust ratio, $\delta_{GDR}$, has been observed to be approximately inversely proportional to the metallicity in galaxies \citep[e.g.,][]{magdis2012,leroy2011}.
\cite{leroy2011} determined the relation between the gas to dust ratio,  and metallicity to be $\log{\delta} = (9.4 \pm 1.1) -(0.85 \pm 0.13) \times (12+\log(\rm{[O/H]}))$.
Using the gas and dust masses from \cite{aravena2016}, $\delta_{GDR} = 41 \pm 13$ in SPT0346-52, giving us $\log{Z/Z_{\odot}} = 0.7 \pm 0.3$, consistent with the \textsc{cloudy} value.

\section{Discussion}
\label{sec:disc}

In this section, we explore various diagnostics of the different phases of the ISM in SPT0346-52.  We begin with dust emission, then move on to line deficits and the electron density.  We then explore the prevalence of ionized versus neutral gas and the non-detection of \oi 63.  Finally, we look into several gas mass estimates for different phases of the ISM.  For reference, Table \ref{table:lines} lists the transitions explored and their excitation properties.
While SPT0346-52 has negligible AGN contribution to its \lfir, it is possible there is a highly dust-obscured AGN that prevents X-ray emission from being observed \citep{ma2016}.
DSFGs can contain AGN; for example, \cite{wang2013} found $17^{+16}_{-6}\%$ of DSFGs in the ALESS (ALMA LABOCA E-CDF-S Submm Survey) sample contain AGN.
SPT0346-52 may also evolve to contain an AGN \citep[e.g.,][]{toft2014}.
We therefore consider both DSFGs and quasar-host galaxies for comparison.

\begin{deluxetable*}{rcccccccc}
\centering
\tablecaption{Targeted Fine-Structure Lines}
\tablenum{4}
\label{table:lines}
\tablehead{\colhead{Line} & \colhead{Transition$^{\mathrm a}$} & \colhead{$\nu_0\ (\rm{GHz})^{\mathrm a}$} & \colhead{$\rm{E_{ion}}^{\mathrm b}$ (eV)} & \colhead{$\rm{T_e}$ (K)$^{\mathrm b}$} & \colhead{$\rm{n_{crit,H}}$ ($\rm{cm^{-3}}$)$^{\mathrm b}$} &  \colhead{$\rm{n_{crit,e^-}}$ ($\rm{cm^{-3}}$)$^{\mathrm a}$} & \colhead{A $(\rm{s^{-1}})^{\mathrm a}$} & \colhead{$g_u^{\mathrm a}$}}
\startdata
\nii 205\micron & $^3$P$_1$-$^3$P$_0$ & 1461 & 14.53 & 70 & $1.76\times 10^2$ & $48$ & $2.1\times 10^{-6}$ & 3\\
\cii 158\micron  & $^2$P$_{3/2}$-$^2$P$_{1/2}$ & 1901 & 11.26 & 91 & $4.93\times 10^1$ & $50$ & $2.1\times 10^{-6}$ & 4\\
\oi 146\micron  & $^3$P$_0$-$^3$P$_1$ & 2060 &  - & 327 & $7.65\times 10^3$ & - & $1.7\times 10^{-5}$ & 1\\
\nii 122\micron  & $^3$P$_2$-$^3$P$_1$ & 2459 & 14.53 & 188 & $3.86\times 10^2$ & $310$ & $7.5 \times 10^{-6}$ & 5\\
\oi 63\micron & $^3$P$_1$-$^3$P$_2$ & 4745 & - & 228 & $3.14\times 10^4$ & - & $9.0\times 10^{-5}$ & 3
\enddata
\tablecomments{The first two columns list the targeted lines and the fine structure transition that emits that line.
$\nu_0$ is the emitted frequency of the line.
$\rm{E_{ion}}$ is the ionization energy needed to remove an electron.
$\rm{T_e}$ is the excitation temperature needed to populate the transition level.
$\rm{n_{crit,H}}$ is the critical density for collisions with hydrogen at $T=100\ \rm{K}$.
$\rm{n_{crit,e^-}}$ is the critical density for collisions with electrons at $T=10 000\ \rm{K}$.
A is the Einstein A coefficient.
$g_u$ is the statistical weight of the upper level.
}
\tablenotetext{a}{\cite{stacey2011}}
\tablenotetext{b}{\cite{cormier2019}}
\end{deluxetable*}

\subsection{Dust Temperatures}

We characterize the FIR continuum and dust temperatures throughout the source-plane using a modified blackbody function.
We use the form from \cite{spilker2016},
\begin{equation}
\label{eq:dust}
S_{\nu_r} \propto (B_{\nu_r}(T_{D})-B_{\nu_r}(T_{CMB}))(1-e^{- \tau_{\nu_r}}).
\end{equation}
$B_{\nu_r}(T)$ is the Planck function at rest-frame frequency $\nu_r$ and temperature $T$.
The blackbody is modified by the dust optical depth, $\tau_{\nu_r}$, which at long wavelengths can be parameterized by $\tau_\nu = (\nu/\nu_0)^\beta = (\lambda_0/\lambda)^\beta$.
In this parameterization, the optical depth reaches unity at wavelength $\lambda_0$, which together with $T_{D}$ determines the peak wavelength and the width of the peak of the dust emission.
The slope of the Rayleigh-Jeans tail of dust emission is $\beta$.
Typically, $\beta \sim 1.5-2$ and $\lambda_0 \sim 100-200$\micron\ in the rest-frame \citep[e.g.,][]{casey2014}.

This modified blackbody is used to fit for the dust temperature in each pixel of the source-plane reconstruction, as well as the global dust temperature.
Five continuum bands (63\micron, 122\micron, 146\micron, 158\micron, and 205\micron) are used to fit the SEDs.
Because of the limited number of photometric points available for these fits, we follow \cite{greve2012} by fixing $\beta = 2.0$ and $\lambda_0 = 100$\micron.
There are two free parameters in each SED fit, the dust temperature, $T_{D}$, and the normalization.

\begin{figure}
\begin{center}
\includegraphics[width=0.4\textwidth]{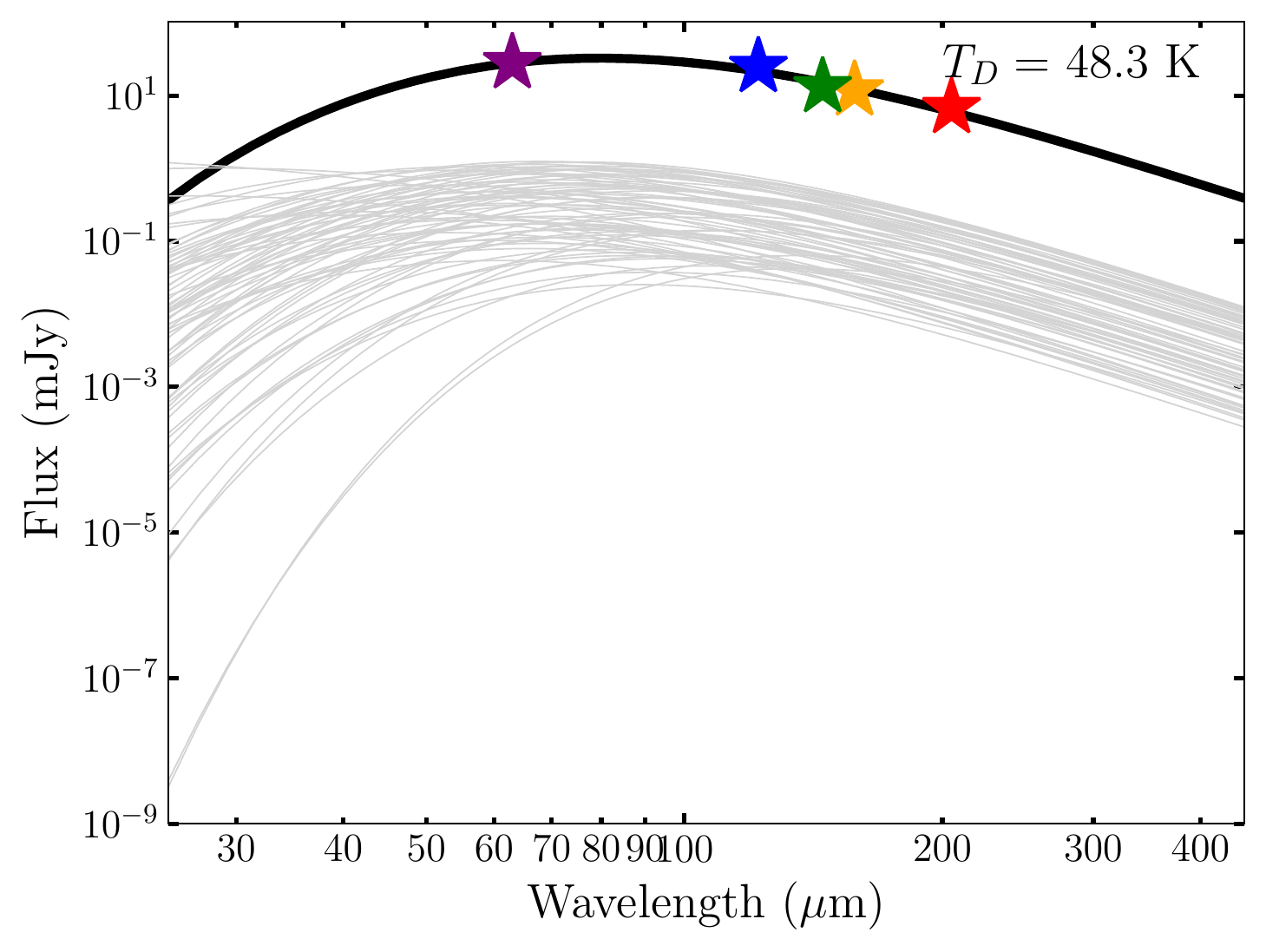}
\includegraphics[width=0.4\textwidth]{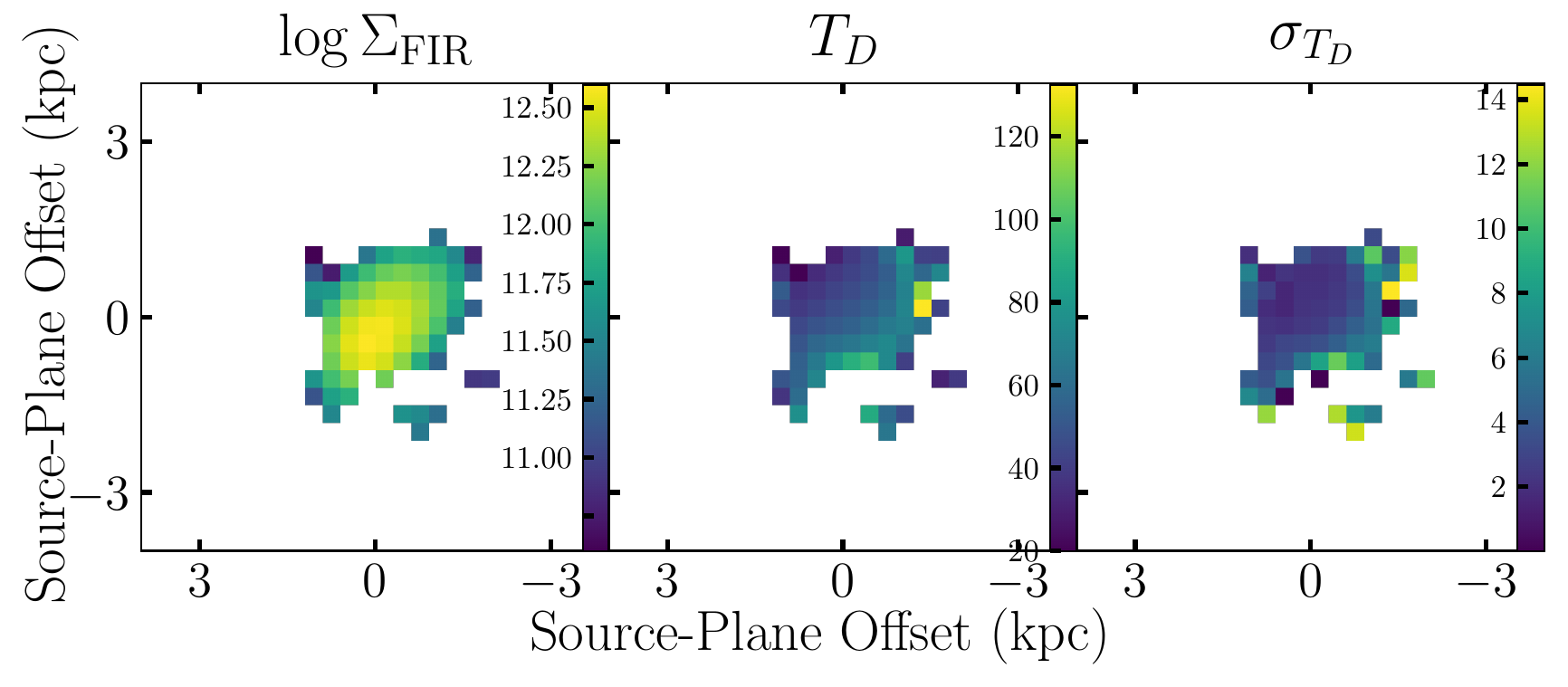}
\caption{Top:  Modified blackbody fits to galaxy-integrated (black) and pixel-by-pixel (grey) continuum flux values. The colored stars are the measured global continuum flux values.  The best-fit SED has a temperature of $T_D = 48.3 K$.  
Bottom left:  $\rm{\Sigma_{FIR}}$, in \lsol/kpc$^{-2}$ across the source.  Bottom center:  Dust temperature throughout the system.  Bottom right:  Error in dust temperature.  The mapped $T_D$ values are the temperatures from the best-fit modified blackbody SEDs in each pixel.  $\rm{\Sigma_{FIR}}$ is calculated by integrating the best-fit modified blackbody SED in each pixel from $42.5-122.5$\micron\ and dividing by the pixel area.
\label{fig:tdustmap}}
\end{center}
\end{figure}

The top panel of Figure \ref{fig:tdustmap} shows the global SED fit (black line) to the continuum (colored stars).  For SPT0346-52, we calculate $T_D = 48.3 \pm 4.0\ \rm{K}$.
The grey lines show the best-fit SEDs for the individual pixels.
The best-fit dust temperatures are shown in the bottom center panel of Figure \ref{fig:tdustmap}, while the right panel shows the error on the best-fit dust temperatures in each pixel.  The left panel shows the FIR surface density ($\rm{\Sigma_{FIR}}$).
\lfir\ is calculated here as the integral of the best-fit modified blackbody SED from $42.5-122.5$\micron\ \citep{helou1988}.
To get $\rm{\Sigma_{FIR}}$, we divide by the source-plane pixel area in $\rm{kpc^2}$.
Dust temperature values in individual pixels have a mean $T_D = 56\ \rm{K}$.

Dust temperatures can vary significantly depending on the fitting form used \citep[e.g., ][]{hayward2012}.
\cite{casey2014} show how, for a given $\lambda_{peak}$, the dust temperature can vary by up to 40 K depending on what assumptions are made about the opacity, as well as the value of $\beta$\ used.
The dust temperatures we calculate are lower than that calculated by \cite{reuter2020} using Equation \ref{eq:dust} ($T_D = 79\pm 15\ \rm{K}$ for SPT0346-52).  However, this temperature difference is a result of the fitting procedures used in this work and by \cite{reuter2020}.  \cite{reuter2020} use the relation between $T_D$ and $\lambda_0$ found by \cite{spilker2016}.  Using this relation tends to increase the fitted dust temperature by $\sim 20\%$ \citep{reuter2020}.
\cite{jones2020} also calculated $T_D = 79\pm 0.5\ \rm{K}$ for SPT0346-52, using a modified blackbody distribution with effects from the CMB.
The small uncertainty is largely due to the unrealistically small photometric errors claimed for the data used in the fit, many of which are well below 1\%.
Using the model from \cite{jones2020} and the data from this paper, we calculate $T_D = 71 \pm 3\ \rm{K}$.
\cite{apostolovski2019} calculated $T_D = 29 \pm 1\ \rm{K}$ for SPT0346-52 by using a radiative transfer model.  
The significant difference from the results of the continuum SED fits presented above undoubtedly results from the difference in methodology and the impact of including CO excitation as a constraint on the dust temperature.

\subsection{Line Deficits}

\begin{figure}
\begin{center}
\includegraphics[width=0.4\textwidth]{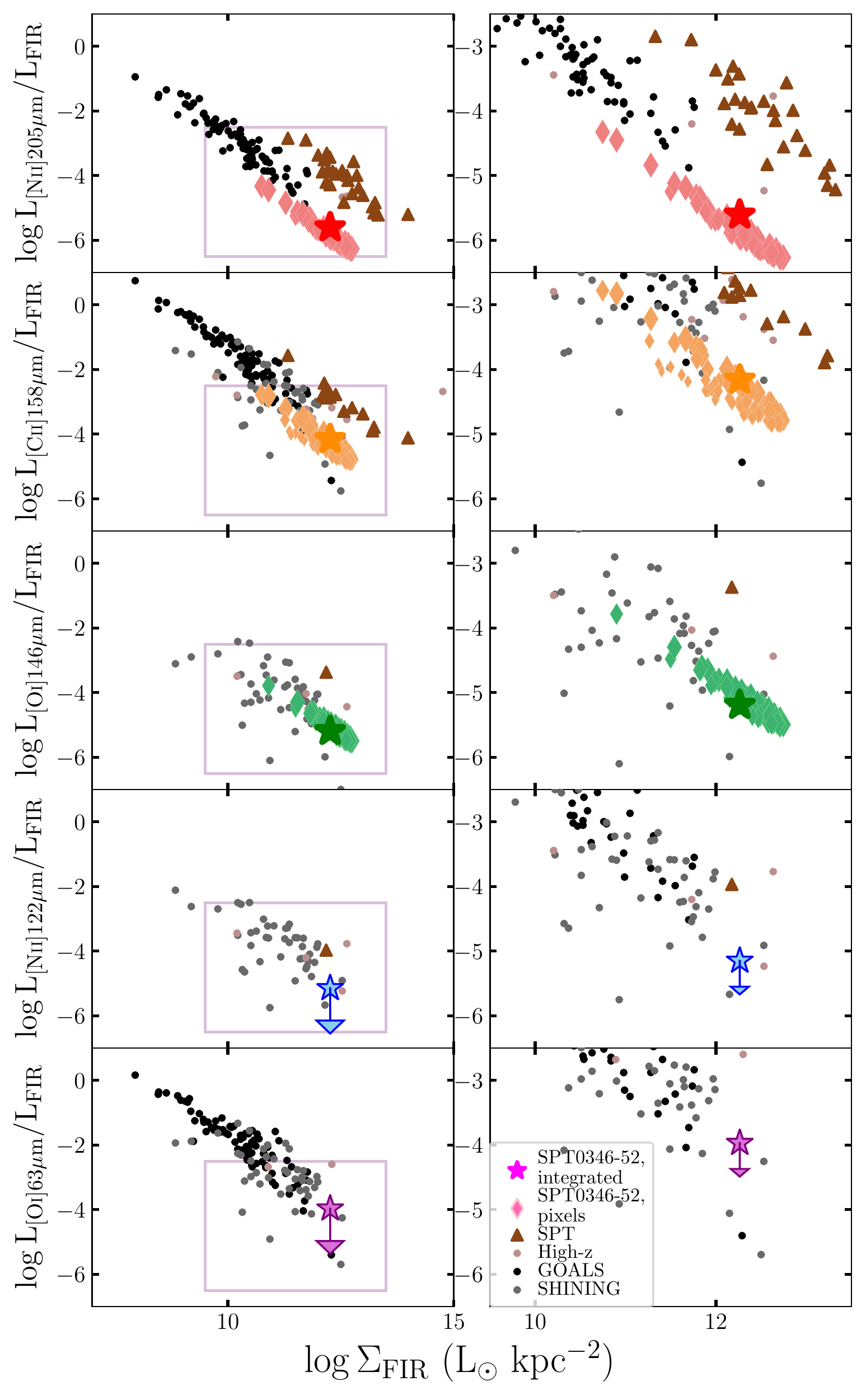}
\caption{Line luminosity/\lfir vs $\rm{\Sigma_{FIR}}$ for the observed lines.  From top to bottom:  \nii 205\micron, \cii 158\micron, \oi 146\micron, \nii 122\micron, \oi 63\micron.The right column shows the same as the left column, but focused on the regions where our pixels lie (the parameter space indicated by light purple boxes in the right column).  Colored diamonds are individual pixels, while the colored stars are the global value for each line. 
For \nii 122 \micron\ and \oi 63\micron, the upper limits are shown.
Brown triangles are galaxies from SPT \citep{gullberg2015,debreuck2019,cunningham2020,reuter2020}.  
Grey dots are galaxy-integrated values from SHINING \citep{herreracamus2018a} and black are from GOALS \citep{diazsantos2017, lutz2016,lu2017}.
Tan dots are a selection of high-z line detections:   J1342+0928 \citep{venemans2017, banados2019, novak2019}, SPT0311-58 \citep{marrone2018}, PJ231-20 \citep{pensabene2021}, HFLS3 \citep{riechers2013}, PJ308-21 \citep{decarli2019,pensabene2021}, G09.83808 \citep{zavala2018,rybak2020}, J2310+1855 \citep{shao2019,li2020}, BR1202-0725 \citep{decarli2014,lu2017,lee2019,lee2021}, SMMJ02399 \citep{weiss2007,ivison2010,ferkinhoff2011}, Cloverleaf \citep{weiss2003,ferkinhoff2011}, SDP.11 \citep{lamarche2018}, and MIPS J1428 \citep{iono2006,haileydunsheath2010}.
\label{fig:deficits}}
\end{center}
\end{figure}

A commonly observed phenomenon is the so-called ``\cii\ deficit'', where $L_{\ciisub}/$\lfir\ falls as $\rm{\Sigma_{FIR}}$ increases \citep[e.g.,][]{luhman1998,sargsyan2012,farrah2013,oteo2016,spilker2016,gullberg2018}.
Here, we explore possible deficits in other FIR lines that trace different components of the ISM.
Figure \ref{fig:deficits} shows the line-to-FIR luminosity ratios versus $\rm{\Sigma_{FIR}}$ for the five lines in this work.
In SPT0346-52, we see deficits, i.e., apparent trends with $\rm{\Sigma_{FIR}}$, in all five lines.
Additionally, we see spatially-resolved deficits (apparent trends with $\rm{\Sigma_{FIR}}$ in individual pixels) in all three detected lines.

\cite{graciacarpio2011} found deficits in \cii 158\micron, \nii 122\micron, \oi 146\micron\ and \oi 63\micron, concluding that line deficits occurred in both ionized and neutral gas in galaxies with a variety of redshifts and optical classifications.  They explained the deficits as resulting from an increase in the ionization parameter  at $\rm{L_{FIR}/M_{H_2}} > 80$\ \lsol/\msol\ as highly compressed, more efficient star formation leads to enhanced ionization parameters.  \cite{zhao2016b} found that there was only a \nii 205 deficit in LIRGs that had warm ($f_{70}/f_{160} > 0.6$) colors.  SPT0346-52 falls into this warm-color regime and does indeed exhibit an \nii 205 deficit.
On the other hand, several individual sources, ranging from $z=1.5$ main sequence galaxies to $z=6$ quasar host galaxies, did not show lower \oi 146\micron/\lfir\ \citep{li2020} and \oi 63\micron/\lfir\ ratios \citep{wagg2020,coppin2012,sturm2010}. 
In the spatially resolved SHINING galaxies, a sample of nearby galaxies that includes star-forming galaxies, AGN host galaxies, and LIRGs, \cite{herreracamus2018a} found the trend of decreasing line/\lfir\ strongest for singly ionized lines like \nii 122\micron\ and \cii 158\micron\ and weakest for neutral \oi 146\micron\ and \oi 63\micron.

\subsection{\cii 158/\nii 205}
\label{sec:ciinii}

Carbon has an ionization potential of 11.26 eV, which is slightly lower than that of hydrogen.  This makes interpretation of \cii 158\micron\  emission difficult because it can originate from both ionized and neutral regions of the ISM.  
By comparing the \cii 158\micron\  to the \nii 205\micron\  emission, which arises only in ionized regions, we can infer how the \cii 158\micron\  emission is divided between neutral and ionized gas.

We can calculate the fraction of \cii 158\micron\  emission originating from neutral gas by comparing the observed \cii 158\micron/\nii 205\micron\  ratio to the expected ratio.  From \cite{abdullah2017},
\begin{equation}
R_{ion} = \frac{I_{\ciisub}}{I_{\niisub}}  = \frac{N_{\ciisub u}}{N_{\niisub u}} \times \frac{E_{\ciisub ul}}{E_{\niisub ul}} \times \frac{A_{\ciisub ul}}{A_{\niisub ul}},
\end{equation}
where $R_{ion}$ is the expected line intensity ratio, $I$ is the expected line intensity of the transition, $N$ is the upper-level population, $E$ is the energy of the transition, and $A$ is the Einstein coefficient for that transition.
This relation assumes the ionic abundance ratio, \cii 158\micron/\nii 205\micron\  is equal to the elemental abundance ratio, $C/N$.
$R_{ion}$ also depends on $n_e$.
As seen in Table \ref{table:lines}, \cii 158\micron\ and \nii 205\micron\ have similar critical electron densities ($50\ \rm{cm^{-3}}$ and $48\ \rm{cm^{-3}}$).  Therefore, if we compute the expected \cii 158\micron/\nii 205\micron\ ratio, there is little density dependence.

\cite{croxall2017} calculated $R_{ion} = 4.0$ using collision rates from \cite{tayal2008} (\cii 158) and \cite{tayal2011} (\nii 205) and assuming Galactic gas-phase abundances for both elements, while \cite{diazsantos2017} used $R_{ion} \simeq 3.0 \pm 0.5$, based on photoionization models by \cite{oberst2006}.  We adopt an intermediate value of $R_{ion} = 3.5$.
With this expected \cii 158\micron/\nii 205\micron\ ratio, we can calculate the fraction of \cii 158\micron\  from neutral gas, $f_{\ciisub, neutral}$, using
\begin{equation}
f_{\ciisub, neutral} = \frac{\ciimath 158 \mu \rm{m} - R_{ion} \times \niimath 205\mu \rm{m}}{\ciimath 158\mu \rm{m}}.
\end{equation}

\begin{figure}
\begin{center}
\includegraphics[width=0.25\textwidth]{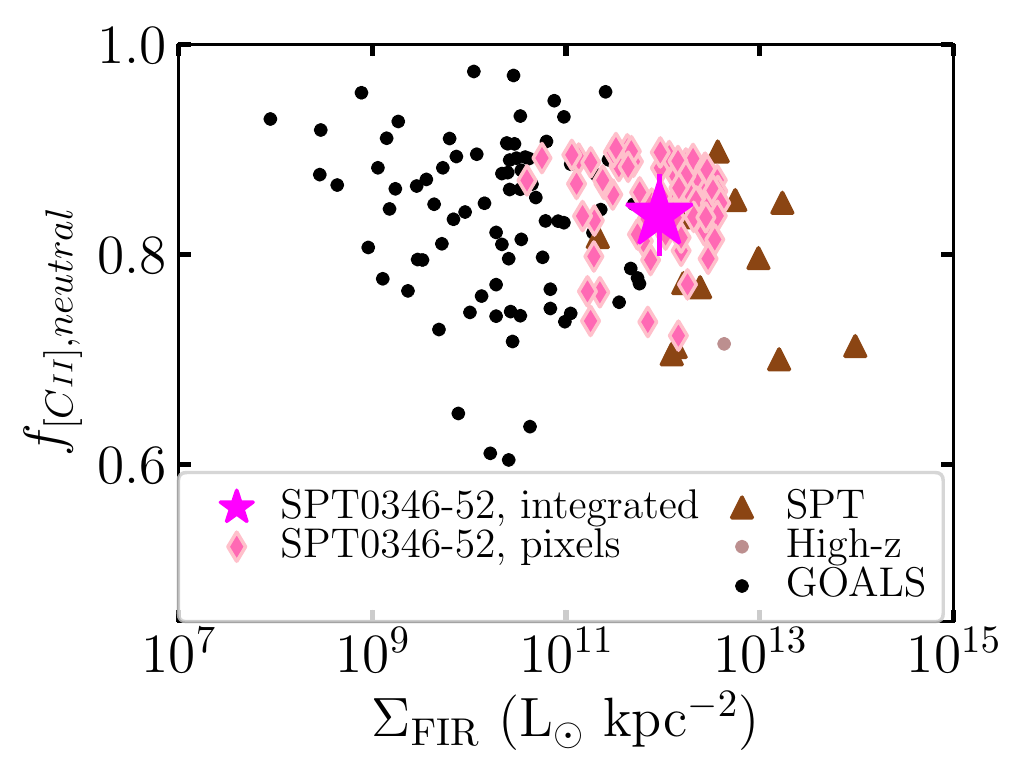}
\includegraphics[width=0.2\textwidth]{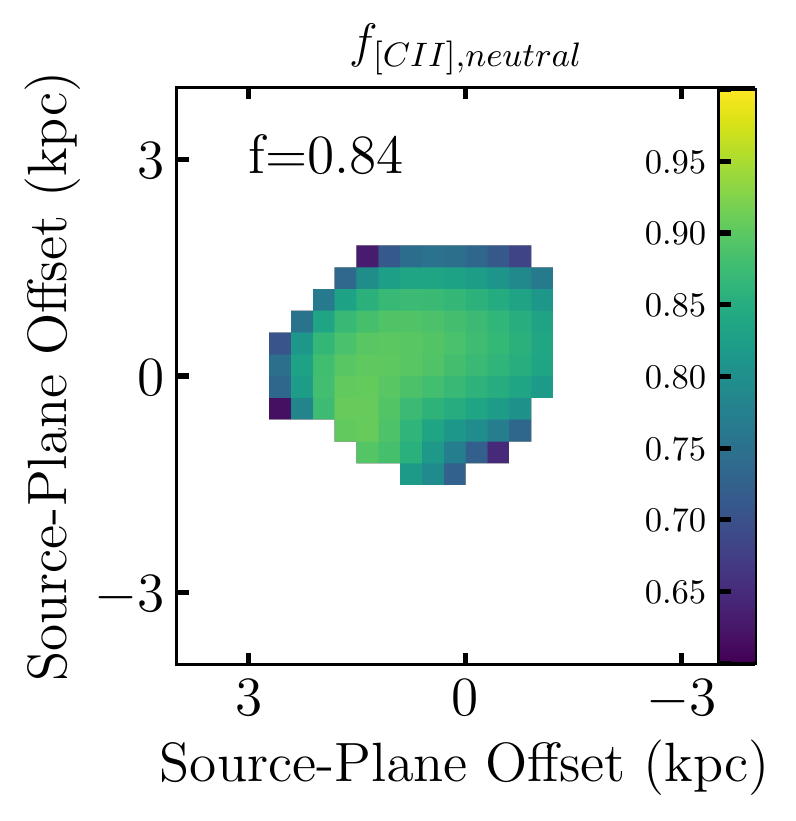}
\caption{Fraction of \cii 158\micron\  emission originating from neutral gas, $f_{\ciisub,neutral}$, calculated from the \cii 158\micron/\nii 205\micron\ ratio.  Left:  SPT0346-52 (pink star) is compared to  galaxies from SPT \citep[brown triangles][]{gullberg2015,cunningham2020,reuter2020}, 
high-z sources \citep[tan dots;][]{decarli2014,lu2017,pensabene2021}, and GOALS \citep[black dots;][]{diazsantos2017, lutz2016,lu2017}.  Right:  Fraction of \cii 158\micron\  originating from neutral gas in SPT0346-52, with the galaxy-integrated value listed in the upper left corner ($f_{\ciisub, neutral} \sim 0.85$).
\label{fig:ciiniimap}}
\end{center}
\end{figure}

For SPT0346-52, we calculate $f_{\ciisub, neutral} = 0.84 \pm 0.04$.
Nearby galaxies ranging from low-metallicity dwarf galaxies \citep{cormier2019} to star-forming galaxies \citep{sutter2019,herreracamus2018a} and (U)LIRGs \citep{diazsantos2017} have $f_{\ciisub, neutral} \sim 60-90\%$ .
As shown in Figure \ref{fig:ciiniimap}, SPT0346-52 has a more comparable $f_{\ciisub, neutral}$ to what has been observed in other high-z sources and DSFGs \citep[$f_{\ciisub, neutral} \sim 85\%$, e.g.,][]{li2020,debreuck2019,pavesi2018,zhang2018}.
This fraction has been observed to be higher in active star-forming regions \citep{herreracamus2018a} and LIRGs with warmer $S_{63}/S_{158}$ colors \citep{diazsantos2017} like SPT0346-52.

\subsection{\cii 158\micron/\oi 146\micron}

\cii 158\micron\ originates from both ionized and neutral gas, while \oi 146\micron\ emission arises from only neutral regions.  
Therefore, more \oi 146\micron\ emission would indicate the presence of more dense, neutral gas.
The ratio of these lines has therefore been used in the literature as an indicator of the prevalence of dense gas \citep{debreuck2019,li2020}.

\begin{figure}
\begin{center}
\includegraphics[width=0.25\textwidth]{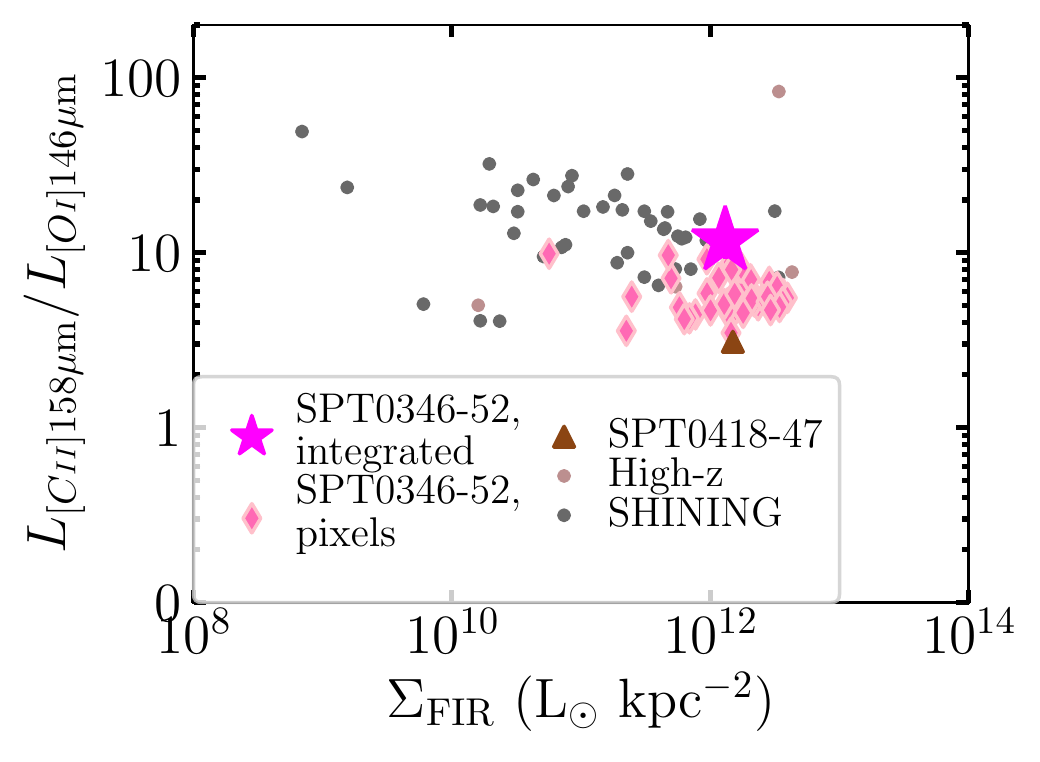}
\includegraphics[width=0.2\textwidth]{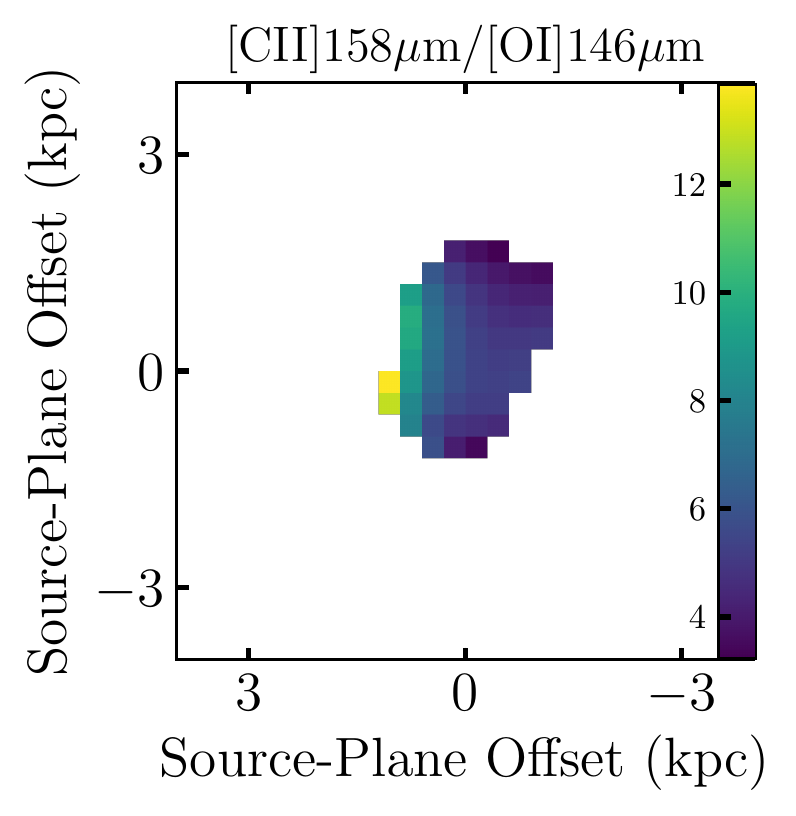}
\caption{\cii 158\micron/\oi 146\micron.
left:  $L_{\ciisub 158 \mu \rm{m}}/L_{\oisub 146 \mu \rm{m}}$ vs $\rm \Sigma_{FIR}$.  Pink diamonds are the individual pixels in SPT0346-52, while 
the pink star represents SPT0346-52.  Comparison samples are taken from \cite{debreuck2019} (SPT0418-47; brown triangle), \cite{lee2021,decarli2014,novak2019} (BR1202-0725 and J1342+0928; tan dots), and \cite{herreracamus2018a} (SHINING; grey dots. 
Right:  Mapped ratio of \cii 158\micron/\oi 146\micron.
Lower $L_{\ciisub 158 \mu \rm{m}}/L_{\oisub 146 \mu \rm{m}}$ values (more \oi 146\micron) indicate more dense, neutral gas.
\label{fig:ciioimap}}
\end{center}
\end{figure}

Figure \ref{fig:ciioimap} plots $L_{\ciisub 158 \mu \rm{m}}/L_{\oisub 146 \mu \rm{m}}$ as a function of $\rm \Sigma_{FIR}$ for SPT0346-52, both integrated and spatially resolved, and values from the literature.  SPT0346-52 has similar \cii 158\micron/\oi 146\micron\ ratio compared to galaxies in the SHINING sample \citep{herreracamus2018a}.
It is higher than SPT0418-47, a $z=4.2$ lensed DSFG.  \cite{debreuck2019} determined that SPT0418-47 had an \cii 158\micron/\oi 146\micron\ ratio $\sim 5\times$ lower than local galaxies, leading them to conclude that the ISM in SPT0418-47 is dominated by dense gas.
\cite{li2020} also found \cii 158\micron/\oi 146\micron\ in their $z\sim 6$ quasar was comparable to the lowest values in ULIRGs, implying that SDSS J2310+1855 has warmer and denser gas compared to local galaxies.
The higher \cii 158\micron/\oi 146\micron\ ratio in SPT0346-52, implying a smaller dense gas component, is consistent with the lower hydrogen gas densities found using \textsc{cloudy} in Section \ref{sec:cloudyresults}.

\subsection{Non-Detection of \nii 122\micron}
\label{sec:nii}

\nii 122\micron\ is expected to be brighter than \nii 205\micron.  However, due to atmospheric O$_2$ at 368.5GHz, just $\sim 800$km/s from the expected center of the \nii 122\micron\ emisssion, \nii 122\micron\ is not detected in SPT0346-52.  
However, we can use the upper limit obtained in Section \ref{sec:obs} to place constraints on the ISM conditions.

Nitrogen ions are only expected to be found in ionized regions of the ISM.  In this regime, the \nii 122\micron\ and \nii 205\micron\  fine structure lines would be excited mostly through collisions with electrons \citep{goldsmith2015}.  Therefore, the relative intensity of \nii 122\micron\ compared to \nii 205\micron\ will depend on the electron density of the ISM and can be used to calculate this density.

\begin{figure}
\begin{center}
\includegraphics[width=0.25\textwidth]{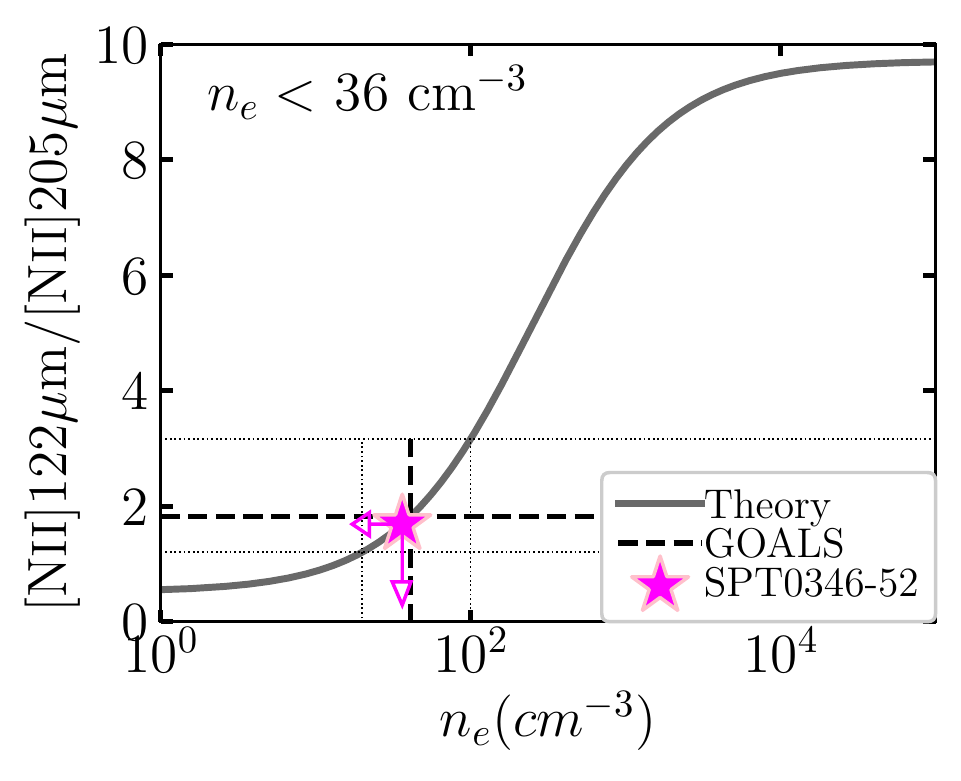}
\caption{Theoretical and calculated electron densities for SPT0346-52.  The grey line is the theoretical relation between the \nii 122\micron/\nii 205\micron\  ratio and electron density. The purple star represents the global \nii 122\micron/\nii 205\micron\  ratio and corresponding $n_e$.  The black dashed line represents the median value from the GOALS sample, and the dotted lines are the maximum and minimum values \citep{diazsantos2017}.
\label{fig:niimap}}
\end{center}
\end{figure}

We calculate the theoretical relation between the \nii 122\micron/\nii 205\micron\ flux ratio and the electron density, $n_e$, following \cite{goldsmith2015} and using the collision rate coefficients from \cite{tayal2011}.  This relation is shown as the grey line in Figure \ref{fig:niimap}.  The left panel also shows the upper-limit to the ratios and corresponding $n_e$ values for SPT0346-52.

In SPT0346-52, we find $n_e < 32\ \rm{cm^{-3}}$.  
The densities observed in SPT0346-52 are lower than the densities calculated in other, comparable systems using the \nii 122/\nii 205 ratio.  For example, \cite{diazsantos2017} found a median $n_e=41\ \rm{cm^{-3}}$ for local LIRGs, with densities ranging from $20\ \rm{cm^{-3}} < n_e < 100\ \rm{cm^{-3}}$ (black dashed and dotted lines in Figure \ref{fig:niimap}), comparable to those found by \cite{zhao2016b}, while \cite{debreuck2019} calculated $n_e \sim 50\ \rm{cm^{-3}}$ for the lensed DSFG SPT0418-47.
The non-detection of \nii 122 indicates low densities in the ionized phase of the ISM.

\subsection{Non-Detection of \oi 63\micron}
\label{sec:oi63}

\cii 158\micron\ and \oi 63\micron\ are both major coolants of the ISM.
As one transitions to high-density ($n > 10^{3-4}\ \rm{cm^{-3}}$), high-temperature ($T > 10^4\ \rm{K}$), high-radiation ($G_0 > 10^3$) regimes, \oi 63 becomes the dominant coolant over \cii 158 \citep{tielens1985,hollenbach1991}.
 One would expect to find bright \oi 63\micron\ emission comparable to or greater than the \cii 158\micron\ emission in FIR-bright systems like starbursts, where the ISM is heated by strong FUV radiation and warmer, denser gas is expected \citep{tielens1985}. 
In the first $z>3$ detection of \oi 63\micron, \cite{rybak2020} found \oi 63\micron\ was $\sim 4$ times brighter than \cii 158 in the $z\sim 6$ DSFG, G09.83808.
Given the intense star formation in SPT0346, we could reasonably expect that \oi63\micron\ might be significantly more luminous than \cii 158\micron. However, we are unable to detect this line in our observations, finding a luminosity ratio of $L_{\ciisub 158}/L_{\oisub 63} > 0.7$. Our inability to place a tighter detection limit reflects the impact of atmospheric O$_2$ absorption centered at 715.4 GHz, just $\sim 1000$ km/s from the redshifted \oi 63\micron\ line center.

The \oi 63\micron\ emission may be intrinsically weak.
\cite{spinoglio1992} also found lower \oi 63\micron\ intensities with their \textsc{cloudy} modeling of starburst regions.
In addition, \cite{abel2009} found that low values of the ionization parameter, $U$, were associated with lower \oi 63\micron/\cii 158\micron\ ratio in their \textsc{cloudy} models of ULIRGs.
For SPT0346-52, \oi 63\micron/\cii 158\micron$ < 1.5$, which is lower than \oi 63\micron/\cii 158\micron$\sim 4$ in the $z=6$ DSFG G09.83808 \citep{rybak2020}.
\cite{rybak2020} found $G=10^4 G_0$ (corresponding to $\log U \approx -2$), which is higher than the values found in SPT0346-52.
Our \textsc{cloudy} modeling of SPT0346-52 indicates lower ionization parameters, consistent with the lower \oi 63\micron/\cii 158\micron\ ratio.

Often \oi 63\micron\  is optically thick 
 \citep{liseau2006,kaufman1999,tielens1985}.
It is also easily self-absorbed; small amounts of cold foreground gas can absorb \oi 63\micron\ while leaving \oi 146\micron\ unaffected \citep{liseau2006}.
This effect has been measured to reduce the \oi 63\micron\ emission by factors of $1.3\pm 1.8$ \citep{kramer2020} up to $2.9\pm 1.6$ \citep{liseau2006}.
If \oi 63\micron/\oi 146\micron\ $< 10$, \oi 63\micron\ is likely self-absorbed \citep{diazsantos2017,cormier2015,tielens1985}.
Taking the $3\sigma$ upper limit and correcting for the lensing magnification, \oi 63\micron/\oi 146\micron$< 14$ in SPT0346-52.
The \oi 63\micron/\oi 146\micron\ intensity ratio may be $< 10$, so \oi 63\micron\ may be self-absorbed.

The \oi 63\micron/\cii 158\micron\ ratio may also be influenced by the presence of an AGN. With its high critical density (Table \ref{table:lines}), \oi 63\micron\ is produced
primarily in dense, neutral gas. In models of PDRs and X-ray dominated regions (XDRs), the \oi 63\micron/\cii 158\micron\ ratio is higher in XDRs (which are expected near AGN) than in PDRs \citep{maloney1996,meijerink2007}.
\cite{ma2016} found no evidence of AGN activity in this source, so we do not expect an AGN-based enhancement of \oi 63\micron\ emission.

\subsection{Gas Mass Estimates}
\label{sec:mass}

\begin{figure}
\begin{center}
\includegraphics[width=0.45\textwidth]{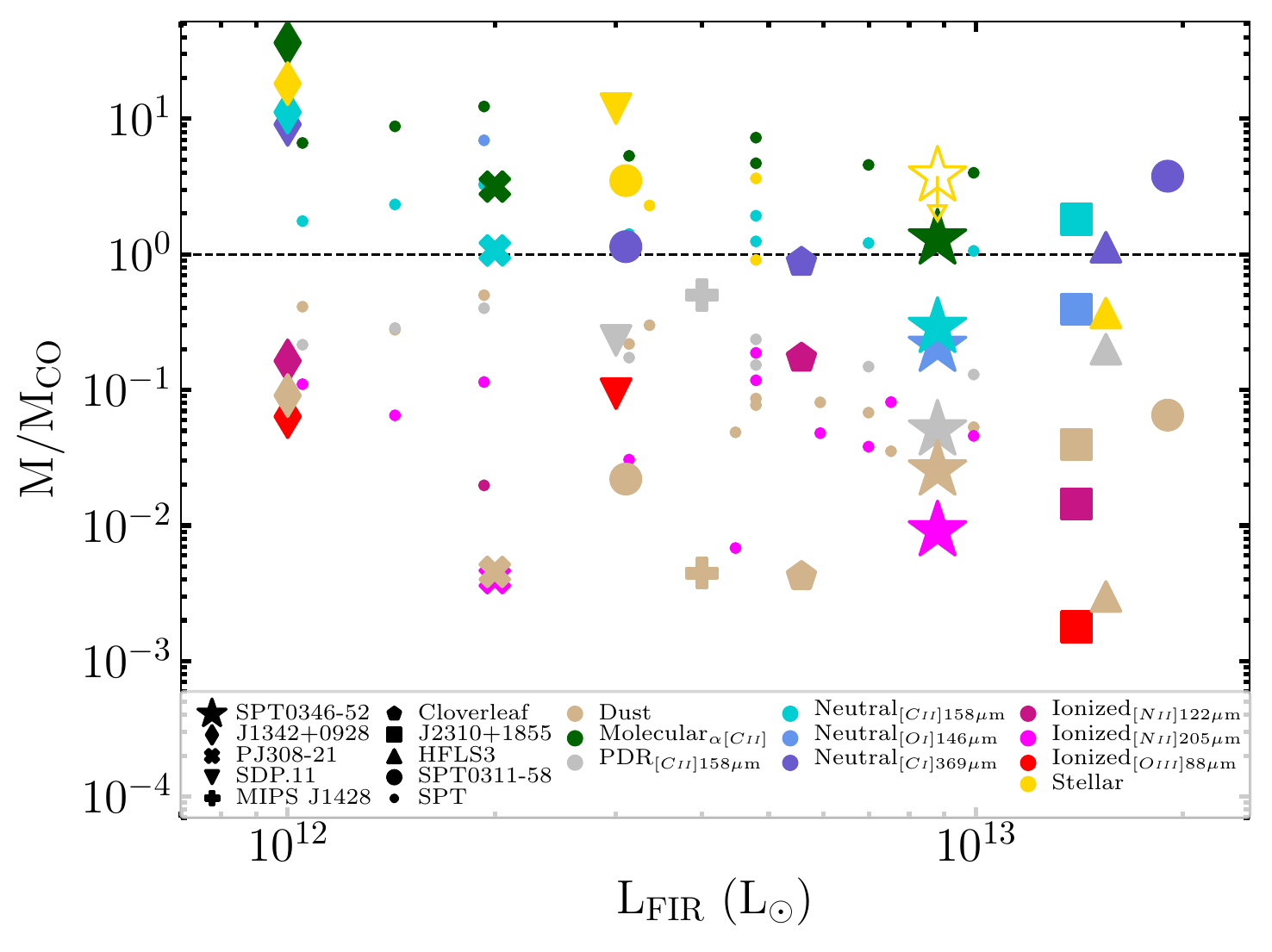}
\caption{Mass estimates for SPT0346-52 and other high-redshift sources, normalized by the molecular gas mass from CO measurements, vs \lfir.  The stars represent the mass estimates for SPT0346-52.  The masses for 
J1342+0928 \citep[$z\approx 7.5$; ][]{venemans2017,novak2019}, 
PJ308-21 \citep[$z\approx 6.2$; ][]{decarli2019,pensabene2021}, 
SDP.11 \citep[$z\approx 1.8$; ][]{lamarche2018}, 
MIPS J1428 \citep[$z\approx 1.3$; ][]{iono2006,haileydunsheath2010}, 
Cloverleaf \citep[$z\approx 2.5$; ][]{weiss2003,ferkinhoff2011},
J2310+1855 \citep[$z\approx 6.0$; ][]{shao2019,li2020}, 
HFLS3 \citep[$z\approx 6.3$; ][]{riechers2013,cooray2014}, and 
SPT0311-58 \citep[$z\approx 7$;]{marrone2018,jarugula2021} are those reported in the literature.
The dust and CO molecular masses for SPT galaxies, including SPT0346-52, are taken from \cite{aravena2016} and \cite{marrone2018} and the stellar masses are taken from \cite{ma2015} and \cite{marrone2018}; the rest of the masses are calculated using the methods described in Section \ref{sec:mass} using data from \cite{gullberg2015,cunningham2020,reuter2020}.
Tan represents dust masses, green is molecular gas estimates using \cii 158\micron\ (Molecular$_{\alpha \ciisub}$) (see Section \ref{sec:molcii}), silver is a PDR model-based gas mass estimate (Section \ref{sec:pdrcii}), blues are neutral gas mass estimates (Section \ref{sec:neutralmass}),  pinks/reds are ionized gas mass estimates (Section \ref{sec:ionmass}), and gold are stellar mass estimates.  The black dashed line represents $\rm{M=M_{CO}}$.
\label{fig:mass}}
\end{center}
\end{figure}

In this section, we estimate the ionized, neutral, and molecular gas masses using the various fine-structure lines observed in SPT0346-52.
For ease of comparison, and because of the higher metallicity expected by gas to dust mass ratio in SPT0346-52,
the ionic abundances are assumed to be the same as the global abundances in \hii\ regions from \cite{savage1996} (i.e., $\chi$(\cii) = C/H, $\chi$(\nii) = N/H, and $\chi$(\oi) = O/H).  
Table \ref{table:masses} lists a summary of the different masses calculated using the various methods described below.  Figure \ref{fig:mass} shows the various masses calculated for SPT0346-52 and several other high-redshift sources, normalized by the molecular gas mass calculated using CO.

\begin{deluxetable}{rcc}
\centering
\tablecaption{Summary of Mass Estimates}
\tablenum{5}
\label{table:masses}
\tablehead{\colhead{Mass Type} & \colhead{Line or Reference} & \colhead{SPT0346-52}\\
\colhead{} & \colhead{} & \colhead{($\times 10^{9}$\ \msol)}}
\startdata
Dust & 1 & $2.1 \pm 0.3$\\
 & 2 & $2.0 \pm 0.6$\\
Molecular & 2 & $82 \pm 6$\\
 & 3 & $390 \pm 220$\\
Stellar & 4 & $<310$\\
\hline
Molecular   & \cii 158\micron & $106 \pm 15$\\
PDR         & \cii 158\micron & $4.1 \pm 0.6$\\
Neutral     & \cii 158\micron & $24 \pm 4$ \\
            & \oi 146\micron & $18 \pm 3$ \\
 Ionized    & \nii 205\micron & $0.8 \pm 0.1$ \\
\enddata
\tablenotetext{1}{\cite{spilker2015}}
\tablenotetext{2}{\cite{aravena2016}}
\tablenotetext{3}{\cite{apostolovski2019}}
\tablenotetext{4}{\cite{ma2015}}
\end{deluxetable}

\subsubsection{Molecular Gas Mass from $\alpha_{\ciisub}$}
\label{sec:molcii}

In their study of $z\sim 2$ main sequence galaxies, \cite{zanella2018} found that the \cii 158\micron\ luminosity and the molecular gas mass of a galaxy were correlated. 
 $\alpha_{\ciisub} = 31 $\msol/\lsol, with a standard deviation of 0.3 dex, is mostly independent of depletion time, metallicity, and redshift.

Using our observed \cii 158\micron\  luminosity, we find a molecular gas mass of $M_{gas,\alpha_{\ciisub}} = 1.1 \pm 0.3 \times 10^{11}$ \msol.  This is between the molecular gas mass calculated by \cite{aravena2016} using CO ($8.2 \pm 0.6 \times 10^{10}$\msol) and the molecular gas mass calculated by \cite{apostolovski2019} using radiative transfer modeling ($3.9 \pm 2.2 \times 10^{11}$\msol).

\cite{zanella2018} calibrated $\alpha_{\ciisub}$ using main sequence galaxies at $z\sim 2$.  However, for DSFGs and quasar host galaxies, this method tends to result in higher molecular gas masses than those obtained using CO, as shown by the green points in Figure \ref{fig:mass}.
The molecular masses calculated using $\alpha_{\ciisub}$ are also higher than the neutral gas masses calculated using the method in Section \ref{sec:neutralmass} in both this work and previous studies \citep{decarli2019, novak2019}.

\subsubsection{Neutral Gas Mass from \cii 158\micron\ PDR Modeling}
\label{sec:pdrcii}

\cite{haileydunsheath2010} calculate the atomic mass associated with PDRs in the $z=1.3$ hyperluminous starburst galaxy MIPS J1428 using the \cii 158\micron\  luminosity and PDR models from \cite{kaufman1999}.  Assuming the \cii 158\micron\  emission is optically thin and that a single temperature characterizes the \cii 158\micron-emitting region, the PDR mass is given by
\begin{multline}
\frac{M_{PDR}}{\rm{M_{\odot}}} = 0.77 \left(\frac{f_{\cii, neutral} L_{\cii}}{\rm{L_{\odot}}}\right)
\left( \frac{1.4\times 10^{-4}}{\chi(C+)}\right) \\ \times \frac{1+2\exp{\left(\frac{-91 K}{T}\right)}+\frac{n_{crit}}{n}}{2\exp{
\left(\frac{-91 K}{T}\right)}}.
\end{multline}
Following \cite{haileydunsheath2010}, we assume the gas temperature is the surface temperature of their modeled PDR, $T \approx 230\ \rm{K}$, $n=10^{4.2}\ \rm{cm^{-3}}$,  $n_{crit} = 2.7\times 10^3\ \rm{cm^{-3}}$ \citep{launay1977}, and $C+/H = 1.4\times 10^{-4}$ \citep{savage1996}.
We use our calculated value of $f_{\ciisub, neutral} = 0.84$.
 We calculate a PDR mass of $M_{PDR} = 4.1 \pm 0.6 \times 10^9$ \msol\ for SPT0346-52.  This is $\sim 5\%$ of the total gas mass from \cite{aravena2016}.

The PDR mass fraction calculated for SPT0346-52 is much lower than that found by \cite{haileydunsheath2010} for MIPS J1428 ($\sim 55\%$).  We also find a similarly lower PDR mass compared to the molecular gas mass in SPT0346-52 than in HFLS3 \citep[$20\%$;][]{riechers2013}, and SDP.11 \citep[$23\%$;][]{lamarche2018}.  
This method assumes a gas density $\log n/cm^{-3} = 4.2$, $\sim 100\times$ higher than the density found using \textsc{cloudy} (Section \ref{sec:cloudyresults}).
If we instead use $\log n/cm^{-3} = 2$, we calculate an atomic PDR mass fraction of $58\%$, which is more comparable to other high-z sources.
This discrepancy could explain the difference in atomic PDR masses between SPT0346-52 and other high-z sources.
Masses calculated with this method are shown as silver points in Figure \ref{fig:mass}.

\subsubsection{Neutral Gas Masses from \cii 158\micron\ and \oi 146\micron}
\label{sec:neutralmass}

We next estimate the neutral gas mass in SPT0346-52 using \cii 158\micron\  and \oi 146\micron.
Based on the work by \cite{weiss2005} and \cite{li2020}, the mass associated with a single transition can be calculated using
\begin{equation}
\label{eq:neutral}
M_{x} = m_x \frac{8 \pi k \nu_0^2}{h c A} \frac{Q(T_{ex})}{g_u} e^{T_e/T_{ex}} L',
\end{equation}
where $m_x$ is the mass of atom $x$, $\nu_0$ is the emission frequency, $A$ is the Einstein coefficient, $Q$ is the partition function, $g_u$ is the statistical weight of the upper level, $k$ is the Boltzmann constant, $h$ is the Planck constant, $T_e$ is the excitation temperature needed to populate the transition level from Table \ref{table:lines}, $T_{ex}$ is the excitation temperature of the gas, and $L'$ is line luminosity in $\rm{K\ km/s/pc^2}$.
The Einstein A coefficients, upper level statistical weights, and transition temperatures can be found in Table \ref{table:lines}.

We adopt an excitation temperature of $T_{ex} = 100\ \rm{K}$.
Based on PDR modeling by \cite{meijerink2007}, $100\ \rm{K}$ is the temperature of a PDR cloud near the outer regions where \cii 158\micron\  is primarily emitted.  The cloud will cool off to $\sim 20\ \rm{K}$ in the inner parts of the cloud as carbon transitions from \cii\ to neutral C to CO.  We also use this temperature for the \oi 146\micron-based mass calculation.
This method assumes the \cii 158\micron\  and \oi 146\micron\ emission is optically thin.

For \cii 158\micron\ , Equation \ref{eq:neutral} becomes
\begin{equation}
\frac{M_{C+}}{\rm{M_{\odot}}} = 3.21\times 10^{-4} \frac{Q(T_{ex})}{4} e^{91/T_{ex}} L'_{\ciisub 158 \mu \rm{m}},
\end{equation}
where $Q(T_{ex}) = 2+4 e^{-91/T_{ex}}$ and $L'_{\ciisub 158 \mu \rm{m}}$ is multiplied by $f_{\ciisub, neutral}$ so only the neutral gas contribution to the \cii 158 emission is included.
  \cite{decarli2019} argue that the mass calculated using \cii 158\micron\  is a lower limit as it does not include non-ionized carbon or effects from lower metallicty gas, suppressed \cii 158\micron\ emission from collisional de-excitation, and optical depth effects.

For \oi 146\micron, we can calculate the oxygen mass using
\begin{equation}
\frac{M_{O}}{\rm{M_{\odot}}} = 6.20\times 10^{-5} Q(T_{ex}) e^{329/T_{ex}} L'_{\oisub 146 \mu \rm{m}}.
\end{equation}
In this case, $Q(T_{ex}) = 5+3 e^{-228/T_{ex}}+e^{-329/T_{ex}}$.

To get the neutral gas mass, we divide by the $C/H$ or $O/H$ abundance.  Using the  abundance from \cite{savage1996} ($C/H = 3.98\times 10^{-4}$ and $O/H = 5.89\times 10^{-4}$), we find neutral gas masses of $M_{neutral, \ciisub 158\mu \rm{m}} = 2.4 \pm 0.4 \times 10^{10}$ \msol\ and $M_{neutral, \oisub 146\mu \rm{m}} = 1.8 \pm 0.3 \times 10^{10}$ \msol.
The \cii 158\micron\  and \oi 146\micron\ neutral gas masses are approximately one-quarter the total gas mass calculated by \cite{aravena2016}.
In general, the neutral gas masses calculated using \cii 158\micron, \oi 146\micron, and \ci 369\micron\ in both SPT0346-52 and in dusty star-forming galaxies and quasar host galaxies in the literature are within a factor of $\sim 10$ of the molecular gas mass (blue points in Figure \ref{fig:mass}).

The mass calculated using \cii 158\micron\ is $\sim 6\times$ higher than the mass calculated using \cii 158\micron\ and PDR modeling in Section \ref{sec:pdrcii}.
While two different temperatures were used (230 K vs 100 K), this difference only changes the calculated masses by $\sim 30\%$ and does not fully account for the large discrepancy in the mass estimates.
The mass estimate in Section \ref{sec:pdrcii} is based on the PDR model of a cloud illuminated on one side from \cite{kaufman1999}.  As discussed in Section \ref{sec:cloudy}, this is a simple model compared to the complexity of a galaxy.
On the other hand, the method used in this section assumes optically thin emission and that the lines are in local thermodynamic equilibrium.  These assumptions may not be valid for all of the \cii 158\micron\ emission in SPT0346-52 and could also account for the discrepancy between the \cii 158\micron\ neutral gas mass estimates.
In addition, as discussed in Section \ref{sec:pdrcii}, the discrepancy also arises from the density of the gas in the PDR model.  Using $n = 10^{2}\ cm^{-3}$ from our \textsc{cloudy} modeling instead of $n = 10^{4.2}\ cm^{-3}$ from \cite{haileydunsheath2010} results in consistent PDR masses using the different methods. 

\subsubsection{Ionized Gas Mass}
\label{sec:ionmass}

Following the method \cite{ferkinhoff2010} used for \oiii 88\micron\ and \cite{ferkinhoff2011} adapted for \nii 122\micron, we can calculate the minimum ionized gas mass required to produce the observed \nii 205\micron\ emission.
This method assumes all the nitrogen in the \hii\ regions is singly ionized, and that the gas is in the high-temperature limit as would be expected in active star-forming regions.
With these assumptions, we can calculate the minimum ionized gas mass using
\begin{equation}
M(\hiimath)_{205} = \frac{L_{\niisub 205 \mu \rm{m}}}{\frac{g_u}{Q(T_{ex})} A_{205} h \nu_{205}} \frac{m_H}{\chi(\niimath)}.
\end{equation}
Here, $h$ is the Planck constant and $m_H$ is the mass of the Hydrogen atom.
In the high temperature limit, the partition function of \nii 205 is $Q(T_{ex}) \rightarrow 9$.  $g_u$, $A$, and $\nu_{205}$ can all be found in Table \ref{table:lines}.
Using the nitrogen abundance from \cite{savage1996}, $N/H = 7.76 \times 10^{-5}$, we find a minimum ionized gas mass of $M(\hiimath)_{205} \le 7.5 \pm 1.0 \times 10^8$ \msol.

\cite{ferkinhoff2011} found that $M_{\hiimath}/M_{mol}$ is correlated with $\rm{\Sigma_{SFR}}$, where more intensely star-forming galaxies have higher fractions of ionized gas.
The minimum ionized gas mass in SPT0346-52 is $\sim 1\%$ of the total molecular gas mass from \cite{aravena2016}.
These values are lower than the ionized gas mass fractions determined for the Cloverleaf \citep[$\sim 8\%$][]{ferkinhoff2011} and J1342+0928 \citep[$\sim 4\%$][]{novak2019}.
As mentioned above, this method assumes the gas is in the high-temperature limit.  This is likely not the case for SPT0346-52.  At lower temperatures, more mass will be required to produce the observed \nii 205\micron\ emission.
The ionized gas masses calculated here are therefore lower limits to the total ionized gas mass.

As shown in Figure \ref{fig:mass}, the molecular phase is the most significant mass component in  SPT0346-52. There is $\sim 4\times$ more molecular gas than neutral gas and  $\sim 100\times$ more molecular gas than ionized gas.
This is in contrast to galaxies in the nearby universe; for example, the Milky Way has a molecular gas mass ($\sim 2\times 10^9$ \msol) very close to its ionized gas mass ($\gtrsim 1.6\times 10^9$ \msol), and more than twice as much atomic gas compared to molecular gas \citep{ferriere2001}.
 The large molecular gas reservoir in SPT0346-52 fuels the large star formation rate observed in this system.

\section{Summary and Conclusions}
\label{sec:conc}

In this work we present ALMA Bands 6, 7, and 9 observations of \nii 205\micron, \cii 158\micron, \oi 146\micron, and undetected  \nii 122\micron\ and \oi 63\micron, as well as the underlying continuum at all five wavelengths, in the $z=5.7$ lensed dusty star-forming galaxy SPT0346-52.  We reconstruct the lensed continuum and line data using the pixelated, interferometric lens modeling code \textsc{ripples} in order to study the source-plane structure of SPT0346-52.  We analyze both the galaxy-integrated properties and the spatially resolved properties of SPT0346-52.

We use the photoionization code \textsc{cloudy} to model the physical conditions of the ISM in SPT0346-52.  
It has lower ionization parameter ($\log{U} \sim -2.75$) and hydrogen density ($\log{n_H/cm^{-3}} \sim 2$) than other high-z DSFGs.
Based on \textsc{cloudy} modeling, we find supersolar metallicity ($\log Z/Z_{\odot} = 0.75$), similar than would be expected from the gas to dust ratio in SPT0346-52.

We calculate the dust temperatures throughout SPT0346 and compare the global dust temperature and the wavelength where the SED peaks to other models.
We look at line deficits and find deficits in all five lines and spatially-resolved deficits in all three detected lines, \nii 205\micron, \cii 158\micron, and \oi 145\micron.
  We use the limit on the \nii 122\micron/\nii 205\micron\ ratio to find $n_e < 32\ \rm{cm^{-3}}$ in SPT0346-52, which is lower than what is observed in ULIRGs and other DSFGs.  Using \cii 158\micron/\nii 205\micron, we determine $\sim 84\%$ of the \cii 158\micron\  emission originates from neutral gas, comparable to other high-z sources and ULIRGs.
Using the \cii 158\micron/\oi 146\micron\ ratio, we see that SPT0346-52 has similar dense gas in PDRs to local galaxies.

Finally, we calculate ionized, neutral, and molecular gas masses using a variety of methods.  The molecular gas mass is $\sim 100\times$ the ionized gas mass and $\sim 4\times$ the neutral atomic gas mass.  The molecular ISM dominates the mass budget of SPT0346-52, fueling the intense star-formation in this system.

\acknowledgments{
We thank the anonymous referee for their insightful and thorough comments.
We also thank Chris Marslender for his computational support.
The SPT is supported by the U.S. National Science Foundation (NSF) through grant OPP-1852617.
K.C.L., D.P.M. J.D.V., K.P. and S.J. acknowledge support from the US NSF under grants AST-1715213 and AST-1716127.
K.C.L and S.J. acknowledge support from the US NSF NRAO under grants SOSPA5-001 and SOSPA4-007, respectively.
J.D.V. acknowledges support from an A. P. Sloan Foundation Fellowship.
M.A. acknowledges partial support from FONDECYT grant 1211951, ANID+PCI+INSTITUTO MAX PLANCK DE ASTRONOMIA MPG 190030, ANID+PCI+REDES 190194 and ANID BASAL project FB210003.
N.S. is a member of the International Max Planck Research School (IMPRS) for Astronomy and Astrophysics at the Universities of Bonn and CologneThe National Radio Astronomy Observatory is a facility of the National Science Foundation 
operated under cooperative agreement by Associated Universities, Inc.
This material has made use of the the Ocelote high-performance computer, which is part of the High Performance Computing (HPC) resources supported by 
the University of Arizona TRIF, UITS, and RDI and maintained by 
the UA Research Technologies department.
This paper makes use of the following ALMA data: ADS/JAO.ALMA
\#2013.1.01231.S, \#2015.1.01580.S, and \#2016.1.01565.S.
ALMA is a partnership of ESO (representing its member states), NSF (USA) and NINS (Japan),
together with NRC (Canada), MOST and ASIAA (Taiwan), and KASI (Republic of Korea),
in cooperation with the Republic of Chile. The Joint ALMA Observatory is operated by
ESO, AUI/NRAO and NAOJ.
This research has made use of NASAÕs Astrophysics Data System.
This research also uses the Cosmology Calculator by \cite{wright2006}.
Cloudy has been supported by NSF (1816537), NASA (ATP 17-ATP17-0141), and STScI (HST-AR- 15018).
}

\software{
 CASA \citep[v5.4.0; ][]{casa,petry2012},
ripples \citep{hezaveh2016},
visilens \citep[\url{https://github.com/jspilker/visilens};][]{hezaveh2013,spilker2016}
 Cloudy \citep[v17.01;][]{ferland2017}, 
FSPS \citep{conroy2009,conroy2010},
 MIST \citep{choi2016,dotter2016}
 }


\bibliographystyle{apj}
\bibliography{dsfgs}

\appendix

\section{Lens Modeling Details}
\label{app:lens}

\begin{figure*}
\begin{center}
\includegraphics[width=0.8\textwidth]{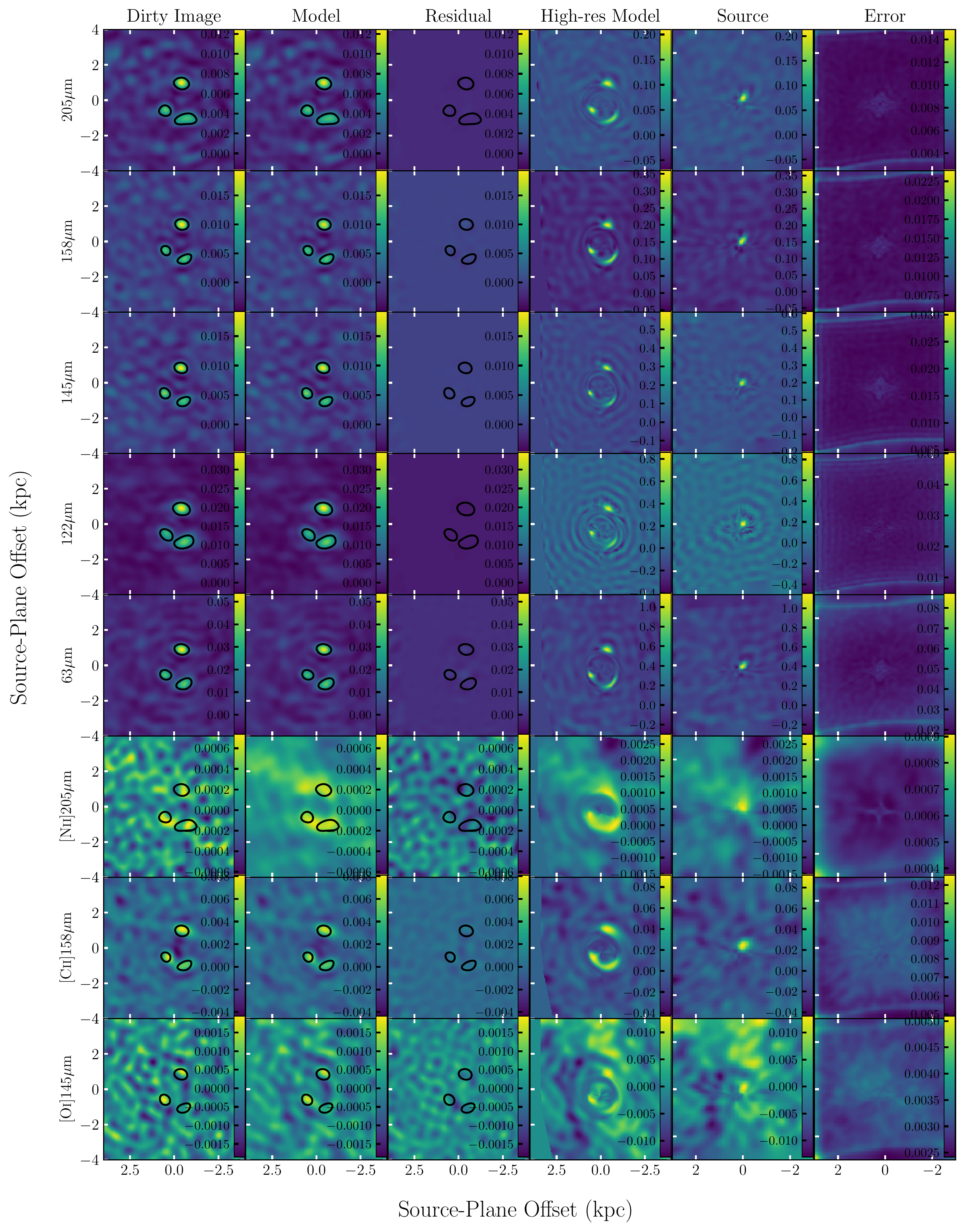}
\caption{
Image-plane and source-plane maps.  From left to right:  dirty image of data visibilities, dirty image of model visibilities, residual map of dirty images, high-resolution (non-visibility sampled) image-plane model, source-plane model, source-plane uncertainty.  From top to bottom:  205\micron, 158\micron, 145\micron, 122\micron, 63\micron, \nii 205\micron, \cii 158\micron, \oi 145\micron.  Contours indicate the observed continuum emission at each wavelength.  The dirty observed image, dirty model image, and residual image are on the same color scale for each set of models.
\label{fig:residual}}
\end{center}
\end{figure*}

Figure \ref{fig:residual} shows the observed data and modeled image- and source-plane data, along with the residual image-plane emission and the uncertainty in the source-plane.  As described in Section \ref{sec:modcomp}, the uncertainty maps are obtained by creating 500 random sets of visibilities and taking the standard deviation in each pixel of the 500 noise reconstructions.

\section{Spatially Resolved Best-Fit \textsc{cloudy} Ratios}
\label{app:cloudy}

\begin{figure*}
\begin{center}
\includegraphics[width=0.9\textwidth]{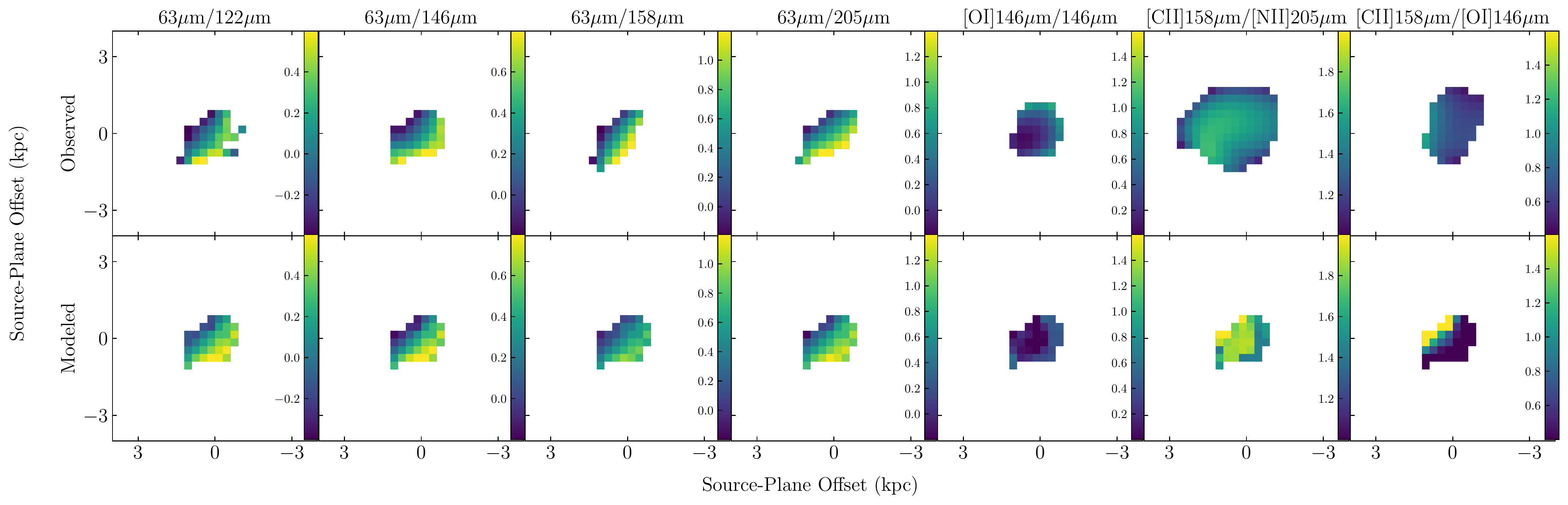}
\caption{Maps of observed and modeled ratios from \textsc{cloudy}.  From left to right:  $\log$ 63\micron/122\micron, $\log$ 63\micron/146\micron, $\log$ 63\micron/158\micron, $\log$ 63\micron/205\micron, $\log$ \oi 146\micron/146\micron, $\log$\cii 158\micron/\nii 205\micron, $\log$\cii 158\micron/\oi 146\micron.  Top:  observed.  Bottom:  modeled.
  Each column has the same color scale.
\label{fig:cloudymaps}}
\end{center}
\end{figure*}

Table \ref{table:ratios} lists the data values compared to the \textsc{cloudy} output to determine for the best-fit models.  For global values, the log of the ratios and the associated errors are listed.
For pixelated values, the mean value is listed.  The recorded error is the mean of the uncertainties on the log of the ratios and represents a typical error in a pixel.  In parentheses, the minimum and maximum pixel values are listed.
Figure \ref{fig:cloudymaps} shows the observed continuum and line ratios, as well as the best-fit model ratios from \textsc{cloudy}.  The fitting procedures and best-fit parameters are described in Section \ref{sec:cloudy}.

\begin{deluxetable}{rcc}
\tablecaption{Data Used to Fit to \textsc{cloudy} Models}
\tablenum{6}
\label{table:ratios}
\tablehead{
\colhead{$\log$ Ratio} & \colhead{Integrated Value$^{\mathrm d}$} & \colhead{Pixel-by-Pixel Values$^{\mathrm e}$}}
\startdata
63\micron/122\micron$^{\mathrm a}$ & $0.05\pm 0.06$ & $0.10\pm 0.07$ (-0.64-0.88)\\
63\micron/146\micron$^{\mathrm a}$ & $0.33\pm 0.06$ & $0.36\pm 0.07$ (-0.14-0.89)\\
63\micron/158\micron$^{\mathrm a}$ & $0.37\pm 0.06$ & $0.55\pm 0.07$ (-0.32-1.35)\\
63\micron/205\micron$^{\mathrm a}$ & $0.62\pm 0.06$ & $0.77\pm 0.07$ (-0.09-1.62)\\
\oi 146\micron/146\micron$^{\mathrm b}$   & $0.38\pm 0.09$ & $0.42 \pm 0.44$ (0.04-0.99)\\
\cii 158\micron/\nii 205\micron$^{\mathrm c}$    & $1.45\pm 0.08$ & $1.46\pm 0.62$ (1.08-1.69)\\
\cii 158\micron/\oi 146\micron$^{\mathrm c}$     & $1.03\pm 0.09$ & $0.71 \pm 0.62$ (0.48-1.11)\\
\enddata
\tablenotetext{a}{Continuum flux density ratio.}
\tablenotetext{b}{Line/continuum ratio in $10^7$ \lsol/mJy (see Section \ref{sec:cloudyparam}).}
\tablenotetext{c}{Line luminosity ratio.}
\tablenotetext{d}{Values and uncertainties used to compare to \textsc{cloudy} models for galaxy-integrated fits.}
\tablenotetext{e}{For each set of pixelated values, we list $\bar{R} \pm \bar{\sigma}$  $(R_{min} - R_{max})$, where $\bar{R}$ is the mean value in the pixels for the log of the ratio, $\bar{\sigma}$ is the mean of the uncertainties in the pixels for the log of the ratio, and $R_{min}$ and $R_{max}$ are the minimum and maximum values of the log of the ratio.}
\end{deluxetable}

\end{document}